\documentclass[12pt]{article}
\usepackage{amsmath}
\usepackage{amssymb}
\usepackage{amsthm}
\usepackage[titletoc,title]{appendix}
\usepackage{graphicx}
\usepackage{float}
\usepackage{wrapfig}
\numberwithin{equation}{section}
\usepackage[left=0.911in,right=0.911in,bottom=1.03in, top=1.03in]{geometry}
\usepackage{setspace}
\usepackage [linktocpage]{hyperref}
\begin{document}
\title{\bf Instantons in AdS$_4$ From (anti)Membranes Wrapping $S^7$ To Bose-Fermi Duality in CFT$_3$'s \ }
\author{{\bf M. Naghdi\,$^{a, b}\,$\footnote{E-Mail: m.naghdi@ilam.ac.ir} } \\
$^a$\,\textit{Department of Physics, Faculty of Basic Sciences}, \\
\textit{University of Ilam, Ilam, Iran}  \\
$^b$\,\textit{School of Particles and Accelerators}, \\
\textit{Institute for Research in Fundamental Sciences (IPM)},\\
\textit{P.O.Box 19395-5531, Tehran, Iran}}
\date{\today}
 \setlength{\topmargin}{-0.25in}
 \setlength{\textheight}{9.25in}
  \maketitle
  \vspace{-0.27in}
    \thispagestyle{empty}
 \begin{center}
   \textbf{Abstract}
 \end{center}
We present new $SO(4)$-invariant and non-supersymmetric instanton solutions for the conformally coupled $m^2=-2$ and massive $m^2=+4$ (pseudo)scalars arising from a consistent truncation of 11-dimensional supergravity over $AdS_4 \times S^7/Z_k$ when the internal space is a $S^1$ Hopf fibration on $CP^3$, and we consider backreaction. In fact, the bulk configurations associate with (anti)membranes wrapped around mixed internal (and external) directions, which in turn probe the Wick-rotated or skew-whiffed background, break all supersymmetries as well as parity invariance. From near the boundary behavior of the closed solution for the coupled bulk (pseudo)scalar, we get a marginal triple-trace deformation with mixed boundary condition (valid also for the bulk massless $m^2=0$ (pseudo)scalar, raised when considering the external space backreaction, with Dirichlet boundary condition) and as a result, the corresponding boundary effective potential is unbounded from below and causes an instability because of the Fubini-like instanton. Presenting dual effective actions, we see that the boundary solutions and counterparts realize in singlet sectors of three-dimensional $U(N)$ and $O(N)$ Chern-Simons- matter field theories. In particular, we use versions of massless and mass-deformed regular and critical boson and fermion models, find instantons and confirm state-operator AdS$_4$/CFT$_3$ correspondence and also Bose-Fermi duality at the level of the solutions. In addition, we discuss on relations of our setups with Vasiliev's Higher-Spin theories, deformations of the Aharony-Bergman-Jafferis-Maldacena model and other related studies.

\newpage
\setlength{\topmargin}{-0.7in}
\pagenumbering{arabic} 
\setcounter{page}{2} 


\section{Introduction}
Instantons, as solutions to Euclidean equations of motion (EoMs) with finite actions, have interesting implications in many physical situations, mainly because of their non-perturbative nature and tunneling features, from quantum corrections to classical behaviors of physical systems to early universe cosmological phenomena. In a few recent studies- see for instance \cite{Me3}, \cite{Me4}, \cite{Me5}, \cite{Me6}- we have tried to find such objects in truncated models of 11-dimensional (11D) and 10D type IIA supergravities (SUGRAs) over $AdS_4 \times S^7/Z_k$ when the internal space (seven-sphere) is considered as an $U(1)$ bundle on $CP^3$, and their dual counterparts in 3D boundary Chern-Simons (CS) superconformal field theories (CFT$_3$) with matter, especially in the  Aharony-Bergman-Jafferis-Maldacena (ABJM) model \cite{ABJM}.

Following \cite{Me7} and \cite{Me8}, in this study \footnote{It should be stressed that this is in fact a study in the structure and not the application, employing models of AdS$_4$/CFT$_3$ correspondence and instanton solutions.}, we employ a 4-form ansatz of the 11D SUGRA, and after solving the original equations, we arrive at truncated scalar equations in 4D Euclidean Anti-de Sitter space ($EAdS_4$). A more interesting case is when one of the (pseudo)scalars in the ansatz, associated with (anti)M2-branes wrapped around three internal directions, is Higgs-like, provides spontaneous symmetry breaking, and the associated solutions can be instantons or bounces corresponding to tunneling or bubble nucleation. Indeed, after counting for the backreaction, that is considering Einstein's equations and then, setting to zero the external and internal components of the energy-momentum (EM) tensors, noting that the topological objects should not change the background geometry, we finally get the equations, which are just for the massless $m^2=0$ and the well-known tachyonic $m^2=-2$ bulk modes, respectively.

In particular, we get a special exact solution for the resultant equation of the latter conformally coupled (CC) (pseudo)scalar, and compute correction to the background 11D SUGRA action because of the solution, and see that it is finite and nonzero, confirming its instanton nature. Then, after taking the bulk solution behavior near the boundary of $AdS_4$, we see that with mixed boundary condition (BC), it corresponds to a marginal triple-trace deformation of the dual boundary CFT$_3$ and that, the corresponding boundary effective potential is unbounded from below and there is an instability associated with the instanton. In addition, we note that the latter marginal deformation may also stand for the massless bulk mode,  which arises when we consider the backreaction of the external space, with Dirichlet BC.

In addition, we write some solutions for the main scalar equation in the bulk, which is valid in probe approximation ignoring the backreaction, for two proper modes of CC and $m^2=+4$ on the skew-whiffed (SW) background. Especially, the solutions of the Nonlinear Partial Differential Equation (NPDE) for the latter (massive) mode will be useful when discussing the boundary duals for the former (CC) mode. More precisely, we see that a double-trace deformation of the boundary dimension-2 operator corresponding to the bulk CC state with Dirichlet BC, in fact corresponds to a single-trace deformation of the resultant dimension-4 operator dual to the massive sate under the Dirichlet BC as well.

On the other hand, because the associated (anti)membranes wrap around the mixed internal and external directions, with respect to (wrt) the ansatz's structure and the equation (\ref{eq104}) used to compute supersymmetries (SUSYs) by counting the killing spinors, the solutions break all original supersymmetries $\mathcal{N}=8 \rightarrow 0$. In addition, because of the mass term in the resultant bulk scalar equations and their high nonlinearity, the scale invariance (SI) is also violated and as a result, the solutions always preserve the $SO(4)$ symmetry of the original isometry group $SO(4,1)$ of $EAdS_4$; while the ansatz and solutions respect the symmetry group $SO(8)\rightarrow SU(4)\times U(1)$ of $S^7 \rightarrow CP^3 \ltimes S^1$ and so, dual boundary operators should be singlets of the same group representations (reps); and for this purpose, we swap the three reps $\textbf{8}_s$, $\textbf{8}_c$ and $\textbf{8}_v$ of $SO(8)$ and then, the needed singlet scalars and pseudoscalars realize after the branching. Meanwhile, the parity invariance is violated with our bulk setups- see also \cite{DuffNilssonPope84}- because of the probe (anti)membranes and so, we should just keep one part of the quiver gauge group $U(N)_k \times U(N)_{-k}$ of ABJM.

On the other hand, we notice that the ABJM model in large $k$ goes to the 3D $O(N)$ vector model; see \cite{CrapsHertog}. In particular, wrt the links between the bulk side of the ABJM \cite{ABJM} and Vasiliev's Higher-Spin (HS) theories in $AdS_4$ ($HS_4$) \cite{Vasiliev01} - see also \cite{Giombi2016} - from one side, and 3D $O(N)$ vector models as duals to $HS_4$ models (see \cite{Klebanov-Polyakov2002} and \cite{Sezgin-Sundell2002} as original studies) from the other side-see also \cite{Chang2013}- and also wrt the recently conjectured Bose-Fermi (BF) duality among 3D $O(N)$ and $U(N)$ CS matter theories (see, for instance, \cite{Aharony2018} and \cite{Choudhury2018}), we try to find boundary solutions dual to the bulk ones in Regular Boson (RB), Regular Fermion (RF), Critical Boson (CB) and Critical Fermion (CF) models in three dimensions. As a result, we find $SO(4)$-invariant instanton solutions in singlet sectors of the massless and mass deformed versions of the models, just keeping one scalar or one fermion with the $U(1)$ (or $SO(2)$) sector of the gauge group and then, test the state-operator (SO) correspondences and confirm the BF dualities with the solutions as well.

This paper is organized as follows. In section 2, after presenting the 11D SUGRA background, 4-form ansatz and then solving the field equation, we get an NPDE scalar equation in $EAdS_4$ for a Higgs-like (pseudo)scalar. Depending on whether the background is Wick-rotated (WR) or SW, we can have a tower of massless, massive and tachyonic (pseudo)scalars realized in different ways in the main scalar equation. Then, we take backreaction into account through computing the EM tensors of the Einstein's equation, and after zeroing them and solving the resultant scalar equations with the main one, we arrive at solvable scalar equations for the massless and CC (pseudo)scalars, from taking the backreaction on the external and internal spaces, respectively. After that, in section 3, we first give an exact solution for the equation of $m^2=-2$ when including the backreaction and then, compute (with details in Appendix C) the instanton correction to the background action in subsection 3.1; and next, in subsection 3.2 (with details in Appendix D) present some perturbative solutions for the latter mode as well as $m^2=+4$ in the SW version of the main equation, as the resulting solutions are useful when performing boundary analyzes. Section 4 is allocated to symmetries and correspondences, where we first review the required AdS$_4$/CFT$_3$ correspondence rules with Neumann, Dirichlet and Mixed BC in Appendix E; and then, in subsection 4.1, we discuss the symmetries of the bulk setups and dual boundary solutions. Afterwards, in section 5, we continue to find boundary counterparts and solutions to the bulk ones; and in this way, we first present the boundary effective actions for the bulk CC (pseudo)scalar in Appendix F, and see that with the Neumann (or mixed) BC corresponding to the operator $\Delta_-=1$, there should be a free or regular scalar (or RB) theory at UV fixed-point (FP) (also dual to type-A HS$_4$ theory with a bulk scalar), while with the Dirichlet BC corresponding to the operator $\Delta_+=2$, the boundary theory should be a free fermion (or RF) theory at IR FP (also dual to type-B HS$_4$ theory with a bulk pseudoscalar); Then, in subsections 5.1 and 5.2, instanton solutions in the massless and mass-deformed versions of the RB and RF models are presented, respectively; In particular, below the subsection 5.1, for the RB (in fact we consider the so-called $\varphi^6$) model, we find a Fubini-like instanton and present interesting physical interpretations. Next, as double-trace deformations with the corresponding operators, take the RB and RF models (at UV) through Renormalization Group (RG) flows to the CB and CF models (at IR), respectively, and that the resulting models are also (duals to bulk arrangements with somehow different BCs from the RB and RF models) needed for dual boundary analyzes, we discuss them briefly and find their instanton solutions in subsections 5.3 and 5.4, respectively; Afterward, in subsection 5.5, we discuss on and confirm the SO correspondence for the bulk modes and dual boundary solutions we have provided; and in subsection 5.6, we confirm the BF duality between RB and CF models and also between RF and CB models, at least at the level of the solutions. Further, in section 6, we discuss on closer connections of our setups and solutions with the Vasiliev's HS theories, ABJM Models and more. Finally, in section 7, we summarize and make a few more comments.

\section{From 11D Supergravity to 4D Gravity}\label{section02}

\subsection{The Background and General Equation}\label{subsection02-01}
We employ the 4-form ansatz
\begin{equation}\label{eq01}
\frac{{G}_4}{(2 R_{AdS})^4} = \frac{3}{8} f_1\, \mathcal{E}_4 - 2\, df_2 \wedge J \wedge e_7 + 8\, {f}_3\, J^2,
\end{equation}
for 11D SUGRA over $AdS_4 \times S^7/Z_k$ as
\begin{equation}\label{eq02}
ds^2_{11D}= R_{AdS}^2\, ds^2_{AdS_4} + R_{7}^2\, \big(ds_{CP^3}^2 + e_7^2 \big), 
\end{equation}
where $S^7/Z_k$ is considered as an $U(1)$ fiber-bundle (with the fiber coordinate $\acute{\varphi}$ in $e_7= (d\acute{\varphi}+\omega)$) on $CP^3$, $R=2 R_{AdS}=R_7$ is the radius of curvature of the bulk $AdS$, $\mathcal{E}_4$ is the unit-volume form on $AdS_4$, $J=d \omega$ is the K\"{a}hler form on $CP^3$ and $f_1, f_2, f_3$ are scalar functions in bulk coordinates. In addition, note that for the ABJM background \cite{ABJM}, the 4-form field reads: $G_4^{(0)}= d\mathcal{A}_3^{(0)} = 3\, R_{AdS}^3\, \mathcal{E}_4 =  N \mathcal{E}_4$, where $N$ is the number of flux quanta on the internal space.

Using the above background and ansatz (\ref{eq01}) - see also \cite{Me7} and \cite{Me8} - we obtain
\begin{equation}\label{eq03a}
df_2 = - 4\, d{f}_3 \Rightarrow {f}_3 = - \frac{1}{4} f_2 \pm {c}_2,
\end{equation}
from the Bianchi identity ($dG_4=0$), and
\begin{equation}\label{eq03b}
 df_1 = i\, 64 R\, f_3\, df_3 \Rightarrow f_1 = i\, 32 R\, f_3^2\, \pm\, i\, {c}_3,
\end{equation}
where the lower/minus sign on the right-hand side (RHS) indicates considering the SW background, and also
\begin{equation}\label{eq03}
 \Box_4 f_3 - \frac{4}{R^2} \left(1 \pm 3\, {C}_3 \right) {f}_3 - 2 \times 192\, {f}_3^3=0,
\end{equation}
from the Euclidean 11D EoM (\ref{eq101}) for the 4-form, respectively, where $\Box_4$ is the $EAdS_4$ Laplacian, and $c_j={C}_j/{R}$ ($j=1,2,3$) with $C_j$'s as real constants.

The interesting case is when we consider the equation (\ref{eq03}) in terms of $f_2$, which reads
\begin{equation}\label{eq06}
     \Box_4 f_2 - M^2\, f_2 \pm \delta\, f_2^2 - \lambda\, f_2^3 \pm F=0,
\end{equation}
where
\begin{equation}\label{eq06a}
M^2=\frac{4}{R^2} \left(1 \pm 3\, C_3 + 288\, C_2^2 \right), \quad \delta = \frac{288}{R}\, C_2,  \quad \lambda=24, \quad F = \frac{16}{R^3}\left(C_2 \pm 3\, C_2\, C_3 + 96\, C_2^3 \right),
\end{equation}
and that the upper signs in phrases like $\pm$, behind the terms including $C_3$, are for the original WR background and the lower signs are for the SW one, while those behind the terms including $C_2$ ($\delta$ and $F$ terms) come from (\ref{eq03a}), noting that picking out the upper or plus sign causes the true vacuum to be on the RHS of the false one (or be in a positive value of $f_2$). Note also that the so-called $\phi^4$ coupling constant (noting $\lambda = 2 \, \lambda_4$) here is $\lambda_4=12$  versus $\lambda_4=192$ in \cite{Me7} (also in (\ref{eq03})) and $\lambda_4=3$ in \cite{Me6}. Now note that depending on the values of $C_3$ and $C_2$, we can have a tower of tachyonic (just for the SW version), massless and massive (pseudo)scalars. However, in this study, we concentrate on the modes $m^2 R_{AdS}^2=0,-2,+4$, and in particular the CC (pseudo)scalar mainly because of its appearance when taking the backreaction in the next subsection as well as other interesting properties when discussing its boundary duals.

It is worth noting that the ansatz (\ref{eq01}) here is indeed a combination of the Freund-Rubin type solutions (nonzero ${G}_4$ just in the external space-time) \cite{FreundRubin1980} and the Englert type solutions (nonzero ${G}_4$ in the internal space) \cite{Englert1982}, with respect to the $SU(4)$ invariant compactification of 11D SUGRA introduced in \cite{Pope1985}. Actually, a more general ansatz and truncation than our (\ref{eq01}), including all bosonic modes, arising from the so-called G-structure tensors (like $e_7$, $J$), are considered in \cite{Gauntlett03}, which in Appendix \ref{Appendix.A22} we face our reduction and setups with corresponding ones in it. It should also be noted that despite the similarities in ansatzs and reductions, in line with our own objectives, we study special bulk modes, equations and solutions as well as various field theory duals, interpretations and other related matters, none of which is discussed in \cite{Gauntlett03}.

\subsection{Taking Backreaction: Massless and CC (pseudo)Scalar Emerge}
To discuss backreaction in the current setup, that is under what conditions the solutions do not backreact on the background geometry (namely there are topological objects), we must first compute the EM tensors (\ref{eq102a}) on the RHS of Einstein equations (\ref{eq102}), and then set them to zero. Doing so, according to the computations done in \cite{Me8} and using (\ref{eq03a}), for the external $AdS_4$ space, the eleventh $S^1/Z_k$ and the internal $CP^3$ components, we gain (noting $f_2\equiv f$ from now on)
\begin{equation}\label{eq08aa}
\Box_4 {f} +\frac{1}{2} \left(- M^2\, f \pm \delta\, f^2 - \lambda\, f^3 \pm F \right)=0,
\end{equation}
\begin{equation}\label{eq08b}
\Box_4 {f} +\frac{1}{4} \left(- M^2\, f \pm \delta\, f^2 - \lambda\, f^3 \pm F \right)=0,
\end{equation}
\begin{equation}\label{eq08c}
\Box_4 {f} + \frac{3}{4} \left(- M^2\, f \pm \delta\, f^2 - \lambda\, f^3 \pm F \right) - \left(-\frac{2}{R^2}\, f \pm \frac{8\, C_2}{R^3}\right) = 0,
\end{equation}
respectively.

Then, we try to solve the latter three equations with the main one (\ref{eq06}), to see under what conditions there is not backreaction on the background geometry. In this way, first note that both (\ref{eq08aa}) and (\ref{eq08b}) are separately satisfied with (\ref{eq06}) giving a trivial constant solution or
\begin{equation}\label{eq08d}
\Box_4 {f} = 0,
\end{equation}
which is the equation for the bulk massless mode, corresponding to an \emph{exactly marginal} boundary operator.\\
In addition, from solving (\ref{eq08c}) with (\ref{eq06}), we obtain
\begin{equation}\label{eq08g}
\Box_4 {f} - \frac{\bar{C}_1}{R^2} \left(-2\, f \pm 8\, c_2 \right) = 0,
\end{equation}
with $\bar{C}_1=4$, corresponding to the well-known CC (pseudo)scalar $m^2 R_{AdS}^2=-2$ \footnote{It is notable that such a mode is in the massless sector of $\mathcal{N}=8$ gauged supergravity in four dimensions, arisen from a consistent truncation of 11D SUGRA over $AdS_4 \times S^7$ \cite{Duff99}, with the "$2/3$" potential
\begin{equation}\label{eq08gga}
   V(\phi)= -3 \cosh\left(\sqrt{\frac{2}{3}} \phi\right)= -3 - \phi^2- ... ,
\end{equation}
where the vacuum or cosmological constant is the first term on the RHS and the second is the mass term.} and \emph{relevant} boundary operators with the bare scaling dimensions $\Delta_{\mp} = 1, 2$ - We return to solve the latter equation and compute the bulk action on its closed solution, in subsection \ref{sub02-03}.

Also, to take the backreaction of the whole internal space, that is solving the equations (\ref{eq08b}), (\ref{eq08c}) and (\ref{eq06}) together, after some mathematical manipulation, we again get (\ref{eq08g}) with $\bar{C}_1=1$ and the scaling dimensions $\Delta_{\pm} = {3}/{2} \pm \sqrt{11}/{2}$. As the same way, to take the backreaction of the whole 11D space, that is to solve the equations (\ref{eq08aa}), (\ref{eq08b}), (\ref{eq08c}) and (\ref{eq06}) together, the result after math manipulations is again (\ref{eq08g}) with $\bar{C}_1=2/3$ and the scaling dimensions
\begin{equation}\label{eq08h}
\Delta_{\pm} = \frac{3}{2} \pm \frac{\sqrt{93}}{6},
\end{equation}
which, as the former, corresponds to a \emph{marginally irrelevant} boundary operator.

\section{Solutions For the Bulk Scalar Equations} \label{sub03-01}
It is possible to find perturbative or approximate solutions for the main bulk equation (\ref{eq06}) using various solving methods of NPDEs. Nevertheless, here we find a closed solution for the equation (\ref{eq08g}) when taking the backreaction and compute the background action correction because of the solution; and then, get approximate solutions for another realization of the same mode $m^2 R_{AdS}^2=-2$ as well as $m^2 R_{AdS}^2=+4$ in the equation (\ref{eq06}) when ignoring the backreaction, that is in probe approximation.

\subsection{Exact Solutions and Correction to the Background Action} \label{sub02-03}
Fist we remind the well-known solution
\begin{equation}\label{eq08f}
      {f}(u,\vec{u}) = C_{4} + \frac{C_{5}\, u^3}{\left[u^2 + (\vec{u}-\vec{u}_0)^2 \right]^3}.
\end{equation}
for the equation (\ref{eq08d}), where we use the upper-half Poincar$\acute{e}$ metric
\begin{equation}\label{eq27}
 ds^2_{EAdS_4} = \frac{R_{AdS}^2}{u^2} \left(du^2 + dx^2 + dy^2 + dz^2 \right).
\end{equation}
Especially, for the equation (\ref{eq08g}) that emerged when taking the backreaction of the internal space, with the scaling of
\begin{equation}\label{eq42}
 \Box_4\, f = -\frac{2\, u}{R_{AdS}^3}\, g + \frac{u^3}{R_{AdS}^3}\, \left(\partial_i \partial_i + \partial_u \partial_u \right)\, g, \quad f=\frac{u}{R_{AdS}} g,
\end{equation}
a closed solution reads
\begin{equation}\label{eq44a}
f(u,\vec{u}) =\frac{4\, C_2}{R} + \frac{2\, b_0}{R}\, \frac{u}{\left[(u+a_0)^2 + (\vec{u}-\vec{u}_0)^2 \right]},
\end{equation}
in which (also in (\ref{eq08f})) $\vec{u}=(x,y,z)$ and $\vec{u}_0=(b_1,b_2, b_3)$, where $a_0$ and $b_j$'s are modulus of the solution showing size and location of the instanton on the boundary, respectively. \footnote{It is notable that a general solution for the equation (\ref{eq08g}) is in terms of Bessel and Hyperbolic functions. However, we use the solution (\ref{eq44a}) as it is more convenient and straightforward for near the boundary analyzes.}

Then, according to the schematic behavior of a physically permissible solution of this type near the boundary as $f(u, \vec{u}) \approx \alpha(\vec{u})\, u^{\Delta_-} + \beta(\vec{u})\, u^{\Delta_+}$, where $\Delta_-$ and $\Delta_+$ are the smaller and larger roots of $m^2\, R_{AdS}^2= \Delta (\Delta-3)$ in $AdS_4$, after the Taylor series expansion of the solution (\ref{eq44a}) around $u=0$, for the CC mode with $\Delta_{\mp}=1,2$, we have
\begin{equation}\label{eq46}
\alpha(\vec{u}) = \frac{2}{R}\, \frac{b_0}{\left[a_0^2 + (\vec{u}-\vec{u}_0)^2 \right]}, \quad  \beta(\vec{u}) = - \frac{2}{R}\, \frac{2\, b_0\, a_0}{\left[a_0^2 + (\vec{u}-\vec{u}_0)^2 \right]^2},
\end{equation}
and so (with $R_{AdS}=1$)
\begin{equation}\label{eq46a}
\beta =-\frac{2\, a_0}{b_0} \alpha^2,
\end{equation}
which in turn corresponds to a \emph{triple-trace deformation} on the dual boundary 3D field theory to which we will return in section \ref{sec05}, particularly in Appendix \ref{Appendix.A5} and subsection \ref{subsec05-02}.

On the other hand, to compute the correction from the instanton (\ref{eq44a}) to the background action, the procedure is outlined in Appendix \ref{Appendix.A2}. As a result, after integrating on the external space coordinates, the finite part of the action, in the unit 7D internal volume, becomes
\begin{equation}\label{eq50}
\tilde{S}_{11}^{corr.} \approx \frac{1}{5} \frac{b_0^2}{a_0^2} \left(\frac{3\, k^3}{\pi^6\, R^5} \right)^{1/2},
\end{equation}
noting that the instanton \footnote{We note that the associated potential from (\ref{eq08g}) is like an upside-down parabola and so, the Fubini-like instanton tunnels from the top of the potential to an arbitrary state.} of the size $a_0$ now sits at the origin ($\vec{u}_0=0$) of a boundary three-sphere with radius $r$ at infinity ($S^3_\infty$). Note also that the correction, for finite $a_0$ and $b_0$, is small in the validity limit ($N\gg k^5$) of the M-theory description of the model.

\subsection{Solutions in Probe Approximation}\label{sub02-02bb}
Because of the importance of the bulk (pseudo)scalars $m^2=-2,+4$ in AdS$_4$ and dual CFT$_3$ discussions, first we consider two ways of their realizations in (\ref{eq06}), valid in probe approximation, as well. The primary way to realize them is with $C_3=1$ and $C_2=0$ in the SW and WR version of (\ref{eq06}) that, of course, we have already discussed their exact and approximate solutions in \cite{Me6} and \cite{Me7}, respectively.

As another way to realize the so-called \textbf{conformally coupled $m^2=-2$} and \textbf{massive $m^2=+4$} (pseudo)scalars, we use for instance the parameters $ C_3=\frac{13}{12}$ \footnote{There is a compelling reason to consider this measure. Indeed, we may take $(1 - 3\, C_3) \equiv \xi\, \mathcal{R}_4$ and then, according to the arguments in \cite{Hrycyna2017} for the value of the non-minimal coupling $\xi=3/16$, the value $C_3={13}/{12}$ is realized, with $\mathcal{R}_4=-12$ for $EAdS_4$.}, $\delta= \frac{6}{\sqrt{2}}$  and $C_3=1, \delta= 12 \sqrt{3}$, in the SW version of the main equation (\ref{eq06}), respectively. Then, one may try to get perturbative solutions, as we have presented in Appendix \ref{Appendix.A3}.

Especially, as we have outlined in the last paragraph of the Appendix \ref{Appendix.A3}, by using the \emph{self-similar reduction} method to solve the main NPDE (\ref{eq06}), we get a perturbative series solution up to the first-order as
\begin{equation}\label{eq41f}
  f^{(1)}(u,r) = \left[ \hat{C}_{\Delta_-} + \check{C}_{\Delta_-}\, \ln(\frac{r}{u}) \right] \left( \frac{u}{r}\right)^{\Delta_-}
         + \left[ \hat{C}_{\Delta_+} + \check{C}_{\Delta_+}\,\ln(\frac{r}{u}) \right] \left( \frac{u}{r}\right)^{\Delta_+}
\end{equation}
near the boundary, where the pairs of the real constants $\hat{C}_{\Delta_-}$ and $\check{C}_{\Delta_-}$ as well as $\hat{C}_{\Delta_+}$ and $\check{C}_{\Delta_+}$ are related to each other, separately; and that for later considerations, we write (\ref{eq41f}) as $f^{(1)}(u,r) \approx \hat{\alpha}\, u^{\Delta_-} + \hat{\beta}\, u^{\Delta_+}$ as well.

It is also interesting to notice that the latter solution is valid not only for $m^2=+4$ (of the equation (\ref{eq08ee})) with $\Delta_{\mp}=-1,4$ but also for $m^2=-2$ (of the equation (\ref{eq08ggaa})) with $\Delta_{\mp}=1,2$ except that for the latter case, $\check{C}_{\Delta_-}=0$. When discussing the boundary dual solutions for the bulk ones in section \ref{sec05} and particularly subsection \ref{subsec05-06}, we see the relevance of the solution (\ref{eq41f}) for deformations with $\Delta_+=2,4$ operators.

\section{Symmetries and Correspondence} \label{Sec-06a}
For this section and at the beginning, we refer the reader to Appendix \ref{Appendix.A4}, where the essential materials of AdS$_4$/CFT$_3$ duality, underlining different quantizations of the bulk (pseudo)scalars on the corresponding boundary theory, is presented. Then, in the following subsection, we discuss briefly the symmetries of the bulk setups and solutions and their implications for the boundary ones.

\subsection{The Main Dual Symmetries}
We note that because the (anti)M2-branes associated with (\ref{eq01}) wrap around the mixed internal directions of $CP^3 \ltimes S^1/Z_k$ and the resulting modes come from the internal ingredients of the 11D field $A_{MNP}$, they are pseudoscalars (of $f_2$ or $f_3$) breaking all supersymmetries and parity as well; see \cite{DuffNilssonPope84}. Further, the current setup could be considered as anti-membranes probing the original M2-branes or membranes probing the SW background, which is in turn for anti-M2-branes; and then, the resulting theory is for anti-M2-branes, with parity breaking. On the other hand, we note that the ABJM model \cite{ABJM} has even parity and so, to build dual solutions with parity breaking pattern, one way is to keep just one part of the quiver gauge group; and also note that, adding CS terms to the boundary matter (scalar and fermion) $O(N)$ and $U(N)$ models, in general breaks parity. 

We also note that the ansatz (\ref{eq01}) and bulk equations and solutions are $SU(4) \times U(1)$-singlet and so the corresponding boundary operators must have the same R-symmetry. \footnote{Note also that the truncation here is consistent, as the arguments in \cite{Duff1985-2} that states the easiest way to achieve a consistent truncation is to keep just a limited numbers of infinite towers of the states that should be singlet of the internal symmetry group.} To fulfill this, besides the SUSY breaking pattern $\mathcal{N}=8 \rightarrow 0$ of the bulk setup, we have already made use of the $SO(8)$ triality that lets swapping the reps $\textbf{8}_s \leftrightarrow \textbf{8}_c$, which means exchanging the supercharges with fermions while keeping the scalars unchanged, and $\textbf{8}_s \leftrightarrow \textbf{8}_v$, which in turn means exchanging the supercharges with scalars while keeping the fermions unchanged, to go from the left-handed (original) to the right-handed (SW) version of the model and find the needed singlet operators of the boundary fields \footnote{We remind that the massless spectrum of 11D SUGRA over $AdS_4 \times S^7$ includes the graviton ($\textbf{1}$), gravitino ($\textbf{8}_s$), gauge fields ($\textbf{28}$), half-integer spin (fermion) fields ($\textbf{56}_{s}$), scalars ($\textbf{35}_{v}$) with $\Delta_- =1$ and pseudoscalars ($\textbf{35}_{c}$) with $\Delta_+ =2$. Under the branching with ${S^7}/{Z_k} \rightarrow CP^3 \ltimes S^1/Z_k$, the three $SO(8)$ reps for the gravitino become
 \begin{equation}\label{eq117ff}
\textbf{8}_s = \textbf{1}_{-2} \oplus \textbf{1}_{2} \oplus \textbf{6}_{0}, \quad \textbf{8}_c = \textbf{4}_{-1} \oplus \bar{\textbf{4}}_{1}, \quad \textbf{8}_v = \bar{\textbf{4}}_{-1} \oplus \textbf{4}_{1} ;
\end{equation}
and scalars, pseudoscalars and gauge fields decompose as
\begin{equation}\label{eq117gg}
   \begin{split}
   & \ \ \ \ \ \ \ \ \ \ \ \textbf{35}_v =\bar{\textbf{10}}_{-2} \oplus \textbf{10}_{2} \oplus \textbf{15}_{0}, \quad \textbf{35}_c =\textbf{10}_{-2} \oplus \bar{\textbf{10}}_{2} \oplus \textbf{15}_{0}, \\
   & \textbf{35}_{s}=\textbf{1}_{-4} \oplus \textbf{1}_{0} \oplus \bar{\textbf{1}}_{4} \oplus \textbf{6}_{-2} \oplus \textbf{6}_{2} \oplus \acute{\textbf{20}}_{0}, \quad \textbf{28} \rightarrow \textbf{6}_{-2} \oplus \textbf{1}_{0} \oplus \bar{\textbf{6}}_{2} \oplus \textbf{15}_{0},
   \end{split}
\end{equation}
respectively. Now, by the swapping $\textbf{8}_s \rightarrow \textbf{8}_v \Rightarrow \textbf{35}_v \rightarrow \textbf{35}_{s}$, and by the swapping $\textbf{8}_s \rightarrow \textbf{8}_c \Rightarrow \textbf{35}_c \rightarrow \textbf{35}_{s}$; As a result, the desired neutral singlet ($\textbf{1}_{0}$) scalar and pseudoscalar are realized, respectively- see also \cite{Me7} and \cite{Me8}, where we have in general discussed how to build singlet scalars and pseudoscalars after the swappings.}- It should also be mentioned that the skew-whiffing breaks all SUSYs except for $AdS_4 \times S^7$ \cite{Duff1984-3}. We also note that the fundamental fields of ABJM are neutral wrt diagonal $U(1)$ that couples to $A_i^+ \equiv (A_i  + \hat{A}_i)$, while $A_i^- \equiv (A_i  - \hat{A}_i)$ acts as baryonic symmetry, and since our (pseudo)scalars are neutral, we set $A_i^-=0$. \footnote{In fact, we may consider deformations like the scalar fluctuations $Y\rightarrow Y + \delta Y$, and this is related to adding the probe (anti)M5-branes wrapped around the internal $R^3 \times S^3/Z_k$ (or the probe (anti)M2-branes wrapped around the internal directions $J \wedge e_7$), which act as domain-walls interpolating among different vacua \cite{Bena}, take the gauge group to $SU(N+1)_k \times SU(N)_{-k}$ and in the $k\rightarrow\infty$ limit, the gauge fields decouple and just the singlet $U(1)$ part remains.}

On the other hand, we note that while the Laplacian $\Box_4$ respects the full $SO(4,1)$ symmetry of the Euclidean $AdS_4$, the mass terms in general break scale- or dilatation-invariance. Although for the CC (pseudo)scalar the latter is not the case, but the solution (\ref{eq44a}) breaks translation- and scale-invariance explicitly and so, as discussed in \cite{Me7}, the solution is indeed $SO(4)$-invariant or $SO(3,1)$-invariant in Lorentzian signature, where the latter is the isometry of $dS_3$ as well. The same argument is valid for the main equation (\ref{eq06}), in addition that it does not respect the $Z_2$ symmetry because of the terms $\sim f$ and $\sim f^3$ in the action from which the equation emerges and so, both the scale- and parity-invariance are broken. \footnote{Note also that beside the SI breaking, for SUSY breaking boundary operators, the conformal dimensions are not generally protected against quantum corrections- see \cite{Avdeev1993} also for other related discussions; However, here we consider the bare dimensions of the operators.}

\section{Dual Solutions in 3D Chern-Simons-Matter Models} \label{sec05}
In this section, we propose dual solutions to the bulk ones. To this aim, we first present the standard boundary effective actions for the bulk CC (pseudo)scalar in Appendix \ref{Appendix.A5}. Then, write solutions wrt the AdS$_4/$CFT$_3$ correspondence rules with physical interpretations. Next, we deal with other boundary scenarios in 3D $O(N)$ and $U(N)$ CS vector models to realize the boundary duals to the bulk solutions; In this way, we also see confirming the BF duality by resulting solutions through the correspondence.

In fact, our bulk solutions are mainly built on the ABJM background and so, the boundary ones we set up should be for the singlet sector of the $U(N)$ gauge group, where the parity breaking happens as well. However, wrt previous studies on 3D $O(N)$ vector models dual to Vasiliev's HS theories in the bulk (see \cite{Klebanov-Polyakov2002} and \cite{Sezgin-Sundell2002} as original studies) and recently raised Bose-Fermi duality in similar situations (see \cite{Aharony2012a} and \cite{Maldacena-Zhiboedov2012} as original studies), we find and discuss dual boundary solutions.

\subsection{Dual Solutions with Regular Boson Model} \label{subsec05-02}
Corresponding to the Neumann and mixed BCs, if we take the singlet boundary scalar $y= \frac{\varphi}{N} I_{N}$, where $\varphi$ is a scalar function on the boundary and $I_N$ is the unit $N\times N$ matrix \footnote{Although here we discard the $N$ factors, the conventions for them could be adopted from \cite{Witten2}.}, we can have $\alpha= \langle \mathcal{O}_{1} \rangle \sim \texttt{tr}(y \bar{y}) \sim \varphi^2$. Then, following the discussions and adjustments in Appendix \ref{Appendix.A5}, we use the \emph{Regular Boson} (RB) model that we write its mass deformed action, plus a CS term, as
\begin{equation}\label{eq122a}
    {S}_{RB}= S_{CS}^{+} + \int d^3\vec{u}\, \left[ \frac{1}{2} (\partial_i \varphi)^2 + \frac{1}{2} m_b^2\, \varphi^2 - \frac{\lambda_6}{6}\, \left(\varphi^2 \right)^3 \right],
\end{equation}
noting that $g_6 \equiv -\lambda_6$ and $\lambda_6 >0$, and that from the corresponding effective action (\ref{eq118b}) we may assign $m_b^2= \mathcal{R}_3/8$; As well as, the CS action reads from \footnote{We remind that to adjust with the ABJM formalism we have already used, where just the $U(1)$ part of the gauge group is kept, set $A_k^-=0$. In addition, $D_k \Phi =\partial_k \Phi + i A_k^{-} \Phi$, with $\Phi$ for both the scalar $Y=y$ and the fermion $\psi$.}
\begin{equation}\label{eq124a}
     \mathcal{L}_{CS}^{+} = \frac{i k}{4 \pi} \varepsilon^{k ij}\ \texttt{tr} \left(A_i^+ \partial_j A_k^+ + \frac{2i}{3} A_i^+ A_j^+ A_k^+ \right).
\end{equation} 
More precisely, as the bulk arrangement leading to the marginal triple-trace deformation of (\ref{eq46a}) or (\ref{eq119}), preserves the full conformal symmetries at the LO, we discard the mass term in (\ref{eq122a}) and so, a solution for the scalar equation, from the resulting so-called tri-critical $O(N)$ model, reads
\begin{equation}\label{eq123}
   \varphi= \left( \frac{3}{\lambda_6} \right)^{1/4} \left[ \frac{a}{a^2 + (\vec{u}-\vec{u}_0)^2} \right]^{1/2},
\end{equation}
which is a Fubini-like instanton on $R^3$ or $S^3_\infty$ of the size $a$ and location parameters $\vec{u}_0$; see \cite{Fubini1} and \cite{Loran2}. Next, as a basic check of the bulk-boundary correspondence, note from (\ref{eq123}) and (\ref{eq46}) that $\varphi^2=\alpha$ with $a=a_0$ and $b_0= a\, ({3}/{\lambda_6})^{1/2}$ - The latter correspondence was interpreted in \cite{deharo06} for the bulk instanton as the square of the boundary one.

Moreover, to confirm the instanton as a finite-action Euclidean solution to the EoM, we compute the action value, based on the solution (\ref{eq123}), as
\begin{equation}\label{eq136a}
    \tilde{S}_{RB} = \frac{\lambda_6}{3} \int \left[ \texttt{tr}(\varphi {\varphi}^{\dag})^3 \right]\, d^3\vec{u}, \quad \int_0^\infty \frac{r^2}{\left( a^2+r^2 \right)^3}\, dr =\frac{\pi}{16\, a^3} \Rightarrow \tilde{S}_{RB}^{c.}= \sqrt{\frac{3}{\lambda_4}} \frac{\pi^2}{4}, 
\end{equation}
where we have used the boundary spherical coordinates with $S^3_\infty$, and assume $a \geq 0$.

\subsubsection{More Interpretations of the Solution}
According to the discussions in \cite{BarbonRabinovici011}, the $SO(4)$-invariant Euclidean solution (\ref{eq123}) has the \emph{bounce} nature, and might also be called \emph{Fubini bubbles} or (true vacuum bubbles) that could nucleate everywhere inside the false vacuum ($AdS_+$), where the four broken symmetries of $SO(4,1)$ act for transferring the expanding bubble around in the 4D volume; see \cite{SmolkinTurok}. The marginal triple-trace deformation here triggers an instability on $S^3$ (or $dS_3$ in Lorentzian signature); these flows could be considered as Coleman-de Luccia \cite{Coleman1980a} instanton backgrounds, which are in turn non-geometrical impurities or smooth bubble-like backgrounds. In addition, one may say that the Kaluza-Klein vacuum decays into nothing, referring to the Witten's bubbles of nothing \cite{Witten3}; see also \cite{Brown-Dahlen2011} where "nothing" (within the bubble of nothing) is thought as the limit of AdS space in which the curvature length shrinks to zero; or shrinking to zero size, moves the effective potential to negative infinity. It is also argued in \cite{Barbon-Rabinovici2011} that these instantons mediate the decay of the classical state ($\varphi=0$) through nucleation of bubbles with the size $a$ at $t=0$, which thus spread in null trajectories.

There are also interpretations about UV-IR correspondence \cite{BarbonRabinovici011}, where the flows with $\Lambda_o \gg 1$ ($\Lambda_o \sim u^{-1}$ as the energy-scale set by the operator perturbation) evolve to a near IR FP at the bubble center; In other words, there is a flow between an UV FP (CFT$_+$) and a IR FP (CFT$_-$) corresponding to a thin-wall bubble of true vacuum ($AdS_-$) inside the false vacuum ($AdS_+$), with the radius $\sim \log(\Lambda_o)$. Nevertheless, the flows below the threshold $\Lambda_o$ do not reach the IR FP, and for the \emph{thick-wall bubble} with $\Lambda_o \ll 1$, the bubble may just be a small hump around its origin. In all cases, marginal or relevant deformations are turned on around the UV FP. In general, the universe inside the bubble is the  Friedmann-Lema\^{\i}tre-Robertson-Walker (FLRW) one with negative curvature; and that the FLRW crunches start at the boundary and propagate inside the bulk \cite{Harlow2010} and so, the first appearance of the crunch is in the deep UV region of the dual 3D field theory and its properties depends on details of the UV-completion. We also remind that indeed there is an infinite negative energy fall in finite time because of the unbounded potential from below; see also \cite{Barbon-Rabinovici2013} for further findings and interpretations.

Meantime, for discussions on big-crunch singularities in $AdS_4$ cosmologies, see also \cite{Kumar-Vaganov2015} and \cite{BzowskiHertog}, where the dual is considered as a constant mass-deformation of the ABJM model on $dS_3$, and also see \cite{Kumar-Vaganov2018} for discussions on resolution of crunch singularities in AdS duals to the boundary $O(N)$ vector models. In addition, in agreement with and completion of the discussion above, it is argued in \cite{Maldacena010} that the bubble decay geometry, as a part of AdS$_4$, is a domain-wall with $dS_3$ geometry and suits to a field theory with a cut-off; and beyond the thin-wall approximation, the CFT$_3$ is deformed with an irrelevant (or relevant) operator, noting that here we are facing the same types of operators and interpretations.

\subsection{Dual Solutions with Regular Fermion Model}
For the Dirichlet BC with $\sigma = \langle \mathcal{O}_{2} \rangle \sim \texttt{tr}(\psi \bar{\psi})$, when we consider the singlet boundary fermion ($\psi_{\hat{a}}^a = \frac{\delta_{\hat{a}}^a}{N} {\psi}$), the boundary effective action (\ref{eq118a}) is indeed for the free or \emph{Regular Fermion} (RF) model and so, we can write its mass deformed action, plus the CS term of (\ref{eq124a}), as
\begin{equation}\label{eq124} 
    {S}_{RF}= S_{CS}^{+} + \int d^3\vec{u}\, \left[ \texttt{tr} \left(\bar{\psi}\, \gamma^i \partial_i \psi \right) + m_f\, \texttt{tr}(\psi \bar{\psi}) \right].
\end{equation}
A solution for the EoM of ${\psi}^{\dag}$ coming from the action (\ref{eq124}) reads (see \cite{Akdeniz1979} for a similar ansatz)
\begin{equation}\label{eq125}
     \psi= \tilde{b}\, \frac{\left[\tilde{a} + i (\vec{u} - \vec{u}_0). \vec{\gamma} \right]^{\varsigma}}{\left[ \tilde{a}^2 + (\vec{u} - \vec{u}_0)^2 \right]^{3/2}}\, \chi,
\end{equation}
where $\varsigma=1$ and the Euclidean gamma matrices are $\vec{\gamma}=(\sigma_2, \sigma_1, \sigma_3)$ with $\sigma_i^{\dag}=\sigma_i$; and that the equation is satisfied if $m_f \rightarrow \tilde{\alpha}(\vec{u})= \texttt{tr}(\psi\bar{\psi})^{1/2} \equiv \hat{\mathcal{O}}_1$ and so, the deformation $W \approx m_f\, \texttt{tr}(\psi \bar{\psi})$ may be considered as a triple-trace one; Note also that $\chi$ is a constant dimensionless spinor with $\chi^{\dag} \chi=1$. As a result, we have
\begin{equation}\label{eq126}
     \langle \mathcal{O}_{2}^{-} \rangle_{\alpha} \sim \texttt{tr}(\psi \bar{\psi}) = \frac{\tilde{b}^2}{\left[ \tilde{a}^2 + (\vec{u}-\vec{u}_0)^2 \right]^{2}} \sim
     \beta(\vec{u}),
\end{equation}
agreed with (\ref{eq46}) \footnote{It should be stressed that the discussions throughout this section, for the duals of the bulk solution (\ref{eq44a}) when taking the backreaction, are valid nearly with perturbative solutions like (\ref{eq53jjj}) (at least in LO) of the equation (\ref{eq08ggaa}) (that is valid in probe approximation) as another way of realization of that state.} with $a_0 =\tilde{a}$ and $b_0=\tilde{b}=a_0$. \footnote{It is notable that if we use the free theory with $m_f=0$, then solving the equation $i \gamma^i \partial_i \psi =0$ gives the same solution (\ref{eq125}) and interpretation except $\tilde{a}=0$. \label{ftn.19.}}

Moreover, the action value for the mass deformed RF model, which is twice the value of the deformation, based on the solution (\ref{eq125}), reads
\begin{equation}\label{eq137a}
    \tilde{S}_{mRF} = 2 \int \left[ \texttt{tr}({\psi} \bar{{\psi}})^{3/2} \right] d^3\vec{u} \Rightarrow \tilde{S}_{mRF}^{c.}= \frac{27\,\pi^2}{2},
\end{equation}
using the same integral as in (\ref{eq136a}), that is again a finite value showing its instanton nature.

\subsection{Dual Solutions with Critical Boson Model}
We note that there is an RG flow from the boundary theory at UV, corresponding to the Neumann BC with $\mathcal{O}_{1}$ when we consider the singlet boundary scalar, to a boundary theory at IR as a double-trace deformation of the former ($\langle \mathcal{O}_{1}^2 \rangle \sim \texttt{tr}(y \bar{y})^2$) with the boundary effective action of (\ref{eq118b}). In fact, the result at IR is the \emph{Critical Boson} (CB) model that we write its mass deformed action, plus the CS term of (\ref{eq124a}), as
\begin{equation}\label{eq127}
    {S}_{CB} = S_{CS}^{+} + \int d^3\vec{u}\, \left[ \frac{1}{2}(\partial_i \varphi)^2 + \frac{1}{2} m_b^2\, \varphi^2 - \frac{\lambda_4}{4} \left(\varphi^2 \right)^2 \right], 
\end{equation}
which is an extension of the Wilson-Fisher model \cite{Wilson-Fisher1972}. \footnote{It should be noted that one may also add the term $\sim \lambda_6\, \varphi^6$ to the CB Lagrangian of (\ref{eq127}) that of course becomes irrelevant in IR and so, we discard it; see \cite{Giombi2017}.} Although in four dimensions, there in an exact solution for the massless so-called $\varphi^4$ theory- see for instance \cite{Fubini1}, \cite{Lipatov1977}, \cite{brzein1977}, \cite{McKane-Wallace1978} and \cite{Actor1982}- as
\begin{equation}\label{eq127a}
 {\varphi} = \left( \frac{8}{\lambda_4} \right)^{1/2} \left[ \frac{a}{a^2 + (\vec{u}-\vec{u}_0)^2} \right],
\end{equation}
where $\lambda_4>0$, but in three dimensions, there is not such an exact solution even for massless case and so, one may use the solution (\ref{eq127a}) roughly just for $a\rightarrow 0$ or $\lambda_4\rightarrow\infty$. As a result, one should in general employ perturbative methods to earn estimated or constrained solutions.

As a standard way, we can rewrite the EoM, from the action (\ref{eq127}), as \cite{Justin1982}
\begin{equation}\label{eq128a}
 \varphi(r)=\left( \frac{m_b}{\lambda_4} \right)^{1/2} h(m_b\, r), \quad  \left( \nabla^2 -1\right) h(r) +  h(r)^3=0,
\end{equation}
where $\nabla^2 = \partial_i \partial^i$ is the 3D Laplacian; and a solution for its linear part, which is the free massive theory, reads
\begin{equation}\label{eq128b}
 h_c(r) \cong \tilde{c}\, \frac{e^{-r}}{r} \Rightarrow  \varphi_c(r) \cong \frac{\tilde{c}}{\sqrt{m_b\, \lambda_4}} \frac{e^{-m_b\, r}}{r},
\end{equation}
which satisfies the condition $h_c(r\rightarrow \infty)\rightarrow 0$, as we are looking for solutions with finite actions \footnote{We note that, according to \cite{Coleman1978a}, the solutions of the Euclidean scalar equations with the lowest action must be spherically symmetric and so, we consider the arbitrary origin $\vec{u}_0$ (or $x_0$) and $r=|\vec{u}-\vec{u}_0|$ (or $|x-x_0|$).}. Besides, solutions of the type (\ref{eq128b}) have already been studied about constrained instantons in four dimensions \cite{Affleck1}; see also \cite{Nielsen1999}. Still, one may use the solution (\ref{eq128b}) as initial data to get perturbative solutions with the generic structure of $\varphi \sim 1/r$, ensuring that it goes to zero at infinity. As a result, with the latter solution,  we have
\begin{equation}\label{eq127b}
 \langle \mathcal{O}_{2}^{+} \rangle_{\beta} \sim \texttt{tr}(y \bar{y})^2 \sim \alpha^2 \sim \frac{1}{r^4},
\end{equation}
which again confirms the SO correspondence with $a_0=0$ in $\alpha$ of (\ref{eq46}). \footnote{Note that the same prescription can be used for the massive RB model of (\ref{eq122a}), that is for the tri-critical $\varphi^6$ model plus the mass term. In fact, the free massive solution (\ref{eq128b}) can be used again as the initial data in perturbative methods; and that we can write a correspondence like (\ref{eq127b}) as $\langle \mathcal{O}_{3}^{+} \rangle_{\beta} \sim \texttt{tr}(y \bar{y})^3 \sim \alpha^3 \sim {1}/{r^6}$, which may also be referred to as a triple-trace deformation again with $a_0=0$ of (\ref{eq46}).}

On the other hand, we note that for the pure $SU(2)$ Yang-Mills theory, an instanton solution (as $r\rightarrow \infty \Rightarrow F_{ij}\rightarrow0$) reads \cite{Belavin}
\begin{equation}\label{eq130}
 \mathcal{A}_i \approx \eta_{i j}\, \frac{(x-x_0)^j}{a^2+(x-x_0)^2} \Rightarrow \mathcal{F}_{i j} \approx \eta_{i j} \left[\frac{a}{a^2+(\vec{u}-\vec{u}_0)^2}\right]^2,
\end{equation}
where $\eta_{i j}$'s are 't Hooft tensors \cite{Hooft1976}, and $\mathcal{F}= d\mathcal{A}+ i \mathcal{A} \wedge \mathcal{A}$ plays role as the dimension-2 operator, which in turn confirms the correspondence $\texttt{tr}(\mathcal{F})\sim \langle \mathcal{O}_{2} \rangle \sim \alpha^2 \sim \beta$, of course with $a=a_0=b_0$ in (\ref{eq46}). Another important point is that, there is a duality between the $\varphi^4$ models with $SO(4)$-invariant solutions and $SU(2)$ Yang-Mills ones, which applies here as well; see \cite{Alfaro1976}, \cite{Corrigan-Fairlie1977}, \cite{Wilczek1976}, \cite{Cervero1977} as original references. Indeed, with $\mathcal{A}_i=\eta_{i j}\, \partial^j \varphi /\varphi$, the $SU(2)$ gauge field equations reduce to the massless scalar equation of $\nabla^2 \varphi+ \lambda_4\, \varphi^3=0$ and so, one may employ one theory to propose solutions for another theory. 

Moreover, for the mass deformed CB model, we may compute the action value based on the estimated solution (\ref{eq128b}) and- see \cite{Justin1982}-
\begin{equation}\label{eq128c}
    \tilde{S}_{mCB} = \frac{\tilde{A}}{\lambda_4}, \quad \tilde{A} = \frac{1}{3} \int (\partial_i h_c)^2\, d^3\vec{u} = \int h_c(r)^2\, d^3\vec{u} = \frac{1}{4} \int h_c(r)^4\, d^3\vec{u}.
\end{equation}
More precisely, we can write the finite contribution of the first integral above from left as
\begin{equation}\label{eq128d}
 \tilde{A} = \frac{4 \pi}{3} \int_0^\infty \left(\frac{dh_c(r)}{dr} \right)^2\, r^2\,dr \Rightarrow \tilde{S}_{mCB}^{c.}= \frac{6 \pi}{\lambda_4},
\end{equation}
with $\tilde{c} \cong 3$ from \cite{Justin2011}.

\subsection{Dual Solutions with Critical Fermion Model}
There is also the \emph{Critical Fermion} (CF) model, which could be considered as a double-trace deformation as $\langle \mathcal{O}_{2}^2 \rangle \sim \texttt{tr}(\psi \bar{\psi})^2 $, noting that there is an RG flow from the CF model at UV to the RF model (\ref{eq124}) at IR; We can write its mass deformed action, plus the CS term of (\ref{eq124a}), as
\begin{equation}\label{eq131}
    {S}_{CF} = S_{CS}^{+} + \int d^3\vec{u}\, \left[ \texttt{tr} \left(\bar{\psi}\, \gamma^i \partial_i \psi \right) + m_f\, \texttt{tr}(\psi \bar{\psi}) + \frac{\tilde{g}_4}{2}\, \texttt{tr}(\psi \bar{\psi})^2 \right],
\end{equation}
which is indeed an extension of the Gross-Neveu (GN) model \cite{Gross-Neveu1974}. \footnote{One may also add the coupling $\tilde{g}_6\, \texttt{tr}(\psi \bar{\psi})^3$ at the large-$N$, which in turn maps to the coupling $g_6$ under the BF duality; see \cite{Giombi2017}.}

A solution to the EoM (of $\bar{\psi}$) coming from the massless action of (\ref{eq131}) has the same structure as (\ref{eq125}) with
\begin{equation}\label{eq132}
     A = \sqrt{\frac{3\, \tilde{a}^{\dag}}{-\tilde{g}_4}} \left( \tilde{a}^2 + (\vec{u} - \vec{u}_0)^2 \right)^{1/2} \Rightarrow \psi \equiv \hat{\psi} = \sqrt{\frac{3\, \tilde{a}^{\dag}}{-\tilde{g}_4}} \frac{1}{\left[\tilde{a}^{\dag}- i  (x - x_0)_k \gamma^{k \dag} \right]}\, \chi .
\end{equation}

It is also fair to say that the solution
\begin{equation}\label{eq134}
     \psi = \pm\, \tilde{B}\, \frac{1}{\left[- i  (x - x_0)_k \gamma^{k \dag} \right]^{1/2}}\, \chi.
\end{equation}
is for the mass deformed action of (\ref{eq131}) and also for its massless version including the coupling $\frac{\tilde{g}_6}{3}\, \texttt{tr}(\psi \bar{\psi})^3$, with $\tilde{B}= \sqrt{\frac{m_f}{-\tilde{g}_4}}$ and $\tilde{B}= \sqrt{\frac{-\tilde{g}_4}{\tilde{g}_6}}$, respectively.

Moreover, the action value for the massless CF model, based on the solution (\ref{eq132}), reads
\begin{equation}\label{eq136b}
    \tilde{S}_{CF} = \frac{3}{2}\, \tilde{g}_4 \int  \left[ \texttt{tr}(\hat{\psi} \bar{\hat{\psi}})^2 \right] d^3\vec{u},   \quad \int_0^\infty \frac{r^2}{\left( \tilde{a}^2+r^2 \right)^2}\, dr =\frac{\pi}{4\, \tilde{a}} \Rightarrow \tilde{S}_{CF}^{c.}= \frac{\tilde{a}}{\tilde{g}_4} \frac{27\,\pi^2}{2},
\end{equation}
assuming $\tilde{a}\geq 0$, and that its finite value again confirms that the solution is indeed instanton.

\subsection{On State-Operator Correspondence} \label{subsec05-06}
To review the SO correspondence with the regular and critical boson and fermion models and near the boundary behavior of the bulk solution (\ref{eq46}), we first note to the RB model with the solution (\ref{eq123}), where the $\Delta_-=1$ operator $ \langle \mathcal{O}_{1}^{+} \rangle \sim \texttt{tr}(y \bar{y}) \sim \alpha$ and so, for the marginal triple-trace deformation, corresponding to the Neumann or mixed BC, $\langle \mathcal{O}_{1}^3 \rangle \sim \texttt{tr}(y \bar{y})^3 \sim \alpha^3$ \footnote{We should note that for any positive value of $g_6$, the operator $\sim g_6\, (\varphi^2)^3$ is quantum \emph{irrelevant}; see \cite{Bardeen1984} and \cite{Papadimitriou2006}; and this statement may roughly match with our bulk solution when taking the backreaction on the whole 11D space, which was associated with a marginally irrelevant deformation (\ref{eq08h}) of the boundary theory. In other words, with quantum corrections, the conformal invariance is no longer exact and so, the exactly marginal configuration may change to a marginally irrelevant one.}. Second, for the RF model, corresponding to the Dirichlet BC, with the solution (\ref{eq125}), we have the $\Delta_+=2$ operator $\langle \mathcal{O}_{2}^{-} \rangle \sim \texttt{tr}(\psi \bar{\psi}) \sim \beta$. Third, for the CB model, corresponding to a double-trace deformation of the RB model with an RG flow from UV to IR, with the LO solution $\sim 1/r$ from (\ref{eq128b}), we have $\langle \mathcal{O}_{1}^2 \rangle \sim \texttt{tr}(y \bar{y})^2 \sim \alpha^2$; Or, one could consider the agreeing solution (\ref{eq130}) and so, $\langle \mathcal{O}_{1}^2 \rangle \sim \texttt{tr}(F) \sim \alpha^2$($\sim \beta$). Fourth, we have the CF model, corresponding to a double-trace deformation of the RF model, with the solution (\ref{eq132}); The CF model is at UV with a dimension-1 operator, which we may consider as $\langle \hat{O}_1 \rangle \sim \texttt{tr}(\hat{\psi} \bar{\hat{\psi}})\sim \alpha$, and there is a flow to RF model at IR having a dimension-2 operator, which may in turn be considered as $\langle \hat{O}_2 \rangle \sim \texttt{tr}(\hat{\psi} \bar{\hat{\psi}})^2 \sim \alpha^2$, though with $\tilde{a}=a_0$ and $b_0=(3/(-\tilde{g}_4)) \tilde{a}$ of (\ref{eq46}).

On the other hand, we remind that the marginal deformations, mainly because of (\ref{eq08d}) with the solutions (\ref{eq08f}) and (\ref{eq08h}), could be arisen from considering the backreaction on the 4D external and the whole 11D spaces, respectively. At the same time, one may just consider the main equation (\ref{eq06}) in probe approximation, with the massless $m^2=0$ mode realized in its SW version for instance with $C_3=1$ and $C_2=1/12$ and so, the LO solution (\ref{eq08f}), valid also for the linear part of (\ref{eq06}), can be used to make perturbative solutions. Either way, the dual boundary theory could be the RB model of (\ref{eq122a}), where the marginal triple-trace deformation could be considered as a single-trace deformation for the Dirichlet BC with the operator $\langle \mathcal{O}_3 \rangle \sim \texttt{tr}(y \bar{y})^3$ and then, the next to the boundary behavior of the main solution (\ref{eq08f}) agrees to the vev of the dimension-3 operator with a solution like (\ref{eq123}) for the boundary scalar, which behaves as $\sim {1}/{(r^2)^3}$; Or one may consider the instanton at the conformal point $u=a$ that give the perfect correspondence in the case.

We also note to the massless CF model, where we could define the operator $\hat{O}_4 \sim \texttt{tr}(\hat{\psi} \bar{\hat{\psi}})^2$ that might be considered as a single-trace deformation dual to the bulk massive (pseudo)scalar $m^2=+4$ under the Dirichlet BC; and so, with the boundary and the bulk solutions (\ref{eq132}) and (\ref{eq41f}) for $\Delta_+=4$ respectively, we have the correspondence
 \begin{equation}\label{eq140b}
     \langle\hat{O}_4\rangle_{\hat{\alpha}} \sim \texttt{tr}(\hat{\psi} \bar{\hat{\psi}})^2 \sim \hat{\beta} \sim \frac{1}{(r^2)^2},
\end{equation}
with $\hat{C}_4=(3 \tilde{a}/(-\tilde{g}_4))^2$, $\check{C}_{4}=0$ and the three-sphere at infinity: $S^3_\infty$.

As the same way, we note that for the RF model with the solution (\ref{eq125}) with $\varsigma = 1$, a double-trace deformation of the dimension-2 operator $\langle \mathcal{O}_{2}^{-} \rangle \sim \texttt{tr}(\psi \bar{\psi})$, acts as a single-trace dimension-4 operator for the Dirichlet BC, and matches with the next to the boundary behavior of the leading bulk solution (\ref{eq50dddd}) as
\begin{equation}\label{eq140c}
     \langle \mathcal{O}_{4} \rangle_{\alpha} \sim \texttt{tr}(\psi \bar{\psi})^2 \sim \beta^2 \sim \frac{1}{(r^2)^{4}},
\end{equation}
with $\tilde{a}=0$ and $\tilde{b}^4=8/\pi^2$; Or the instanton at the conformal point $u=\tilde{a}$ to match with the original solution (\ref{eq50dddd}) exactly. In addition, we note that if we take the solution (\ref{eq125}) with $\varsigma = 2$, then
\begin{equation}\label{eq140d}
    \langle \hat{O}_{2} \rangle_{\hat{\alpha}} \sim \texttt{tr}(\psi \bar{\psi}) \sim \hat{\beta} \sim \frac{1}{(r^2)},
\end{equation}
where $\hat{\beta}$ is read from (\ref{eq41f}) for $\Delta_+=2$, with $\tilde{a}=0$ and $\tilde{b}^2=\hat{C}_{2}$; and also notice that a double-trace deformation of the latter $\hat{{O}}_{2}^{-}$ has the same structure as in (\ref{eq140b}) except that for the current case, $\tilde{a}=0$ and $\tilde{b}^4=\hat{C}_{4}$ must then be set; and that $\check{C}_{\Delta_+}=0$ for both cases.

\subsection{On Bose-Fermi Duality}
The BF duality or 3D Bosonization in CS matter theories was originally studied in \cite{Aharony2012a} and \cite{Maldacena-Zhiboedov2012}; see also \cite{Minwalla-Yokoyama2015}, \cite{Choudhury2018} and \cite{Aharony2018} for further studies. Here we look for a test of this duality at the level of the solutions and correspondences. To this end, from the solution (\ref{eq132}) of the CF model, on the one hand, we read
\begin{equation}\label{eq133a}
     \texttt{tr}(\hat{\psi} \bar{\hat{\psi}}) = \left( \frac{3}{-\tilde{g}_4} \right) \frac{\tilde{a}}{\left[ \tilde{a}^2 + (\vec{u}-\vec{u}_0)^2 \right]},
\end{equation}
and from the solution (\ref{eq123}) of the RB model, on the other hand,  we read
\begin{equation}\label{eq133b}
     \texttt{tr}(y \bar{y}) = \left( \frac{3}{\lambda_4} \right) \frac{{a}}{\left[{a}^2 + (\vec{u}-\vec{u}_0)^2 \right]},
\end{equation}
which are in agreement having $\tilde{a} = a$ and $-\tilde{g}_4= -{g}_4=\lambda_4$ and so, confirming the BF duality  $\varphi\leftrightarrow \hat{\psi}$ of the RB and CF models. 

As the same way, we can make similar statements for the RF and CB models. Indeed, if we take the solution ($\sim \tilde{c}/r$) from (\ref{eq128b}) for the CB model and the massless solution of (\ref{eq125}) for the RF model (see the footnote \ref{ftn.19.}), we can write $\texttt{tr}(y \bar{y})^2 \sim \texttt{tr}(\psi \bar{\psi}) \sim {\tilde{b}^2}/{r^4}$, with $\tilde{a}=0$ and $\tilde{b}=\tilde{c}^2$ and so, the duality is somehow realized with $\psi \leftrightarrow \varphi^2$. Even more interestingly, we have a little different solution of (\ref{eq125}) with $\varsigma=2$, whose massless version fulfills the exact BF duality $\varphi \leftrightarrow \psi$ as
\begin{equation}\label{eq135b}
     \texttt{tr}(y \bar{y}) \sim \texttt{tr}(\psi \bar{\psi}) \sim \frac{\tilde{b}^2}{r^2},
\end{equation}
again with $\tilde{a}=0$ but $\tilde{b}=\tilde{c}$. Further, if we take the solution (\ref{eq130}) with the gauge field $\mathcal{A} \sim \partial\varphi/\varphi \sim \ln(\varphi)$ or a scalar solution like (\ref{eq127a}) for the CB model on the one hand, and the solution (\ref{eq125}) for the RF model on the other hand, we can write
\begin{equation}\label{eq135}
   \texttt{tr}(\psi \bar{\psi}) \sim \texttt{tr}(\mathcal{A} \wedge \mathcal{A}) \sim \frac{a^2}{\left[{a}^2 + (\vec{u}-\vec{u}_0)^2 \right]^2},
\end{equation}
with $a=\tilde{a}=\tilde{b}$, which confirm the (gauge) Boson-Fermion duality $\psi \sim \mathcal{A}$ as well.

\section{Links to Vasiliev HS - ABJM Models and More}
Taking the boundary 3D $O(N)$ vector models as duals to the Vasiliev's HS theories \cite{Vasiliev01} in $AdS_4$ ($HS_4$), was studied in \cite{Klebanov-Polyakov2002} and \cite{Sezgin-Sundell2002}, originally. Indeed, in the simplest bosonic version of the $HS_4$ model, the spectrum of fluctuations around the vacuum includes the CC (pseudo)scalar as well as a tower of massless fields of all spins $s=1,2,...,\infty$, although there is also a truncation including just even spins; and that, all $HS_4$ models include a spin-2 field (graviton). In particular, for the bulk scalar (even parity) matching to the type-A $HS_4$ model, with the alternate BC $\Delta_-=1$ and the operator $\langle \mathcal{O}_1^+ \rangle \sim \varphi^2$, the dual is the (singlet sector of) 3D $O(N)$ model with N free massless scalars at UV FP; and a double-trace relevant deformation, $\langle \mathcal{O}_2^+ \rangle \sim (\varphi^2)^2$, takes it to the critical scalar $O(N)$ model at IR (the Wilson-Fisher model \cite{Wilson-Fisher1972}). At the same time, for the bulk pseudoscalar (odd parity) matching to the type-B $HS_4$ model, with the standard BC $\Delta_+=2$ and the operator $\langle \mathcal{O}_2^- \rangle \sim \texttt{tr}(\psi \bar{\psi})$, the dual is the (singlet sector of) 3D $O(N)$ model with N free fermions at IR FB; and a double-trace irrelevant deformation, $\langle \mathcal{O}_4^- \rangle \sim \texttt{tr}(\psi \bar{\psi})^2$, takes it to the critical fermion $O(N)$ model at UV FB (the Gross-Neveu model \cite{Gross-Neveu1974}). In other words, the regular BC $\Delta_+=2$ corresponds to the strongly coupled IR FP of the O(N) vector model, and is related to the UV FP with a Legendre transform of the type in (\ref{eq116}). It is also notable that the IR $\leftrightarrow$ UV pattern is always realized through the conformal inversion $x^{\mu}\rightarrow x^{\mu}/x^2$; and that the $u$ direction is for the RG flow, noting that the boundary is at $u=0$ (UV) and the horizon is at $u\rightarrow \infty$ (IR); see also \cite{Girardello2002}, \cite{Petkou2003}, \cite{Leigh-Petkou2003} and \cite{Sezgin-Sundell2003}.

More generally, the parity-invariant type- A and B HS theories are dual to the singlet sectors of the $O(N)$ and $U(N)$ boson and fermion vector models; and introducing the CS terms, with breaking the parity (indeed, according to \cite{Chang2013}, turning on the bulk parity breaking phase of $\theta_0$ corresponds to turning on the CS coupling), was performed in \cite{Aharony2011} and \cite{Giombi2011}, respectively. In other words, we can say that the type-A theory with the Neumann BC is dual to a free (single) boson ($U(N)$ or $O(N)$) vector model, and the type-B theory with the Dirichlet BC is dual to a free (single) fermion vector model plus a CS term at the level of $k$ (as (\ref{eq124a})) for both. \footnote{It is important to limit to the singlet sector of the $O(N)_k$ and $U(N)_k$ models in HS/CFT duality. In addition, when one includes CS terms, one indeed takes the limits $N, k\rightarrow\infty$ and $\lambda=N/k$ fixed and so, the singlet sector for the large-$N$ is realized with $\lambda\rightarrow0$. In other words, with massless fundamental fermions coupled to the $U(N)$ CS gauge fields at the level of $k$, there is a family of interacting CFT$_3$ labeled by $k$ and $N$; and with $k\rightarrow\infty$, one reaches the singlet sector of the free fermion vector model dual to the type-B Vasiliev model. In fact, for these limits, there is a line of non-SUSY CFTs parametrized by $\lambda$; and at $\lambda=0$, there is the free fermion vector model.} It is also notable that, the bosonic models of these types were studied in \cite{Aharony2011} with the operators $\bar{O}_1=\varphi^{\dag} \varphi$ and $\bar{O}_2=\bar{\psi} \psi+ \frac{4 \pi}{k} (\varphi^{\dag} \varphi)^2$ and the vertices like $(\varphi^{\dag} \varphi)^3$ and $(\bar{\psi} \psi) (\varphi^{\dag} \varphi)$. In addition, there are dual discussions on the marginal or triple-trace deformations like $\langle \mathcal{O}_1^+ \rangle ^3 \sim (\varphi^2)^3$ and $\mathcal{O}_3^- = \mathcal{O}_1^+ \, \mathcal{O}_2^- $  in \cite{Chang2013}; see also \cite{Hikida-Wada2017}.

It is also interesting to look at likenesses of the setups here with the exact $SO(3,1)$-invariant solution of the $HS_4$ models in \cite{Sezgin-Sundell2005-01}- see also \cite{Sezgin-Sundell2005-02} and references therein- where the minimal bosonic models are obtained from consistent truncations of HS gauge theories based on the symmetry $shs(8|4)\supset osp(8|4)\supset so(3,2) \times so(8)$. The resultant models include $\textbf{35}_+$ (we call $\textbf{35}_v$) scalars and $\textbf{35}_-$ (we call $\textbf{35}_c$) pseudoscalars  in the supergravity multiplet (at the level of $\ell=0$), and $\textbf{1}_+$ scalar and $\textbf{1}_-$ pseudoscalar in the Konishi multiplet (at the level of $\ell=1$); and the solutions in type-A and type-B make use of the Konishi scalar and pseudoscalar, respectively - Indeed, $\langle \mathcal{O}_1^+ \rangle \sim \texttt{tr}(y \bar{y})$ and $\langle \mathcal{O}_2^- \rangle \sim \texttt{tr}(\psi \bar{\psi})$ could be those non-BPS Konishi operators, which also break SUSYs; see also \cite{Sezgin-Sundell2003}. It is also noticeable that there is a truncation of the $\mathcal{N}=8$ HS theory to $\mathcal{N}=6$ HS theory in $AdS_4$ \cite{Sezgin-Sundell2012} including 32 spin-0 particles; that is, in its supergravity multiplet (at the level of $\ell=0$), there are \textbf{15} and $\bar{\textbf{15}}$ and also $\textbf{1}_+$ and $\textbf{1}_-$ (at the level of $\ell=1$) of irreducible reps of SO(6), for scalars and pseudoscalars respectively. Meanwhile, it is good to look at \cite{Iazeolla-Raeymaekers2015}, where an analogues solution in a 3D HS theory (the so-called " Prokushkin-Vasiliev theory"), with a holographic dual, similar to the solution here, is discussed as well.

In addition, relevant to the discussion here is \cite{Vasiliev2012} where, making use of the \emph{unfolding dynamic} method, it is shown that the nonlinear HS theories in $AdS_4$ are dual to nonlinear 3D gauge theories. As well as, see \cite{Maldacena-Zhiboedov2012}, where employing the slightly broken HS symmetry in $AdS_4$, the well-known 3D free and critical CS fermion and boson $O(N)$ models, even beyond the large-$N$ limit, are recovered; Indeed, the latter reference, besides \cite{Aharony2012a}, is one of the original work proposing the BF duality, in which "quasi-fermion" and "quasi-bosons" are referred to scalars with dimensions two and one, respectively. It is also good to have a look at \cite{Chang2013}, where a triality among type IIA string theory over $AdS_4 \times CP^3$, the Vasiliev’s HS theory and $U(N)_k \times U(M)_{-k}$ ABJ theory \cite{ABJ} is surveyed in details.

We also note that the boundary models we use, could be realized as deformations of the ABJM model. In fact, the RB and CB (RF and CF) models could be realized with keeping just a singlet scalar (fermion) and $U(1)$ part of the quiver gauge group; and that, the singlet scalar (fermion) is realized under the skew-whiffing of $\textbf{8}_s \leftrightarrow \textbf{8}_v$ while $\textbf{8}_c$ fixed ($\textbf{8}_s \leftrightarrow \textbf{8}_c$ while $\textbf{8}_v$ fixed). In addition, the RB and CB (RF and CF) 3D boundary models, realized with the Neumann or mixed (Dirichlet) boundary conditions, correspond to the type-A (type-B) $HS_4$ bulk models. More precisely, the massless or massive RB model (\ref{eq122a}), with a triple-trace deformation of the singlet operator $\mathcal{O}_1^{+}$, could be realized in the ABJM original scalar action (in \cite{Me3}, \cite{Me4}) or its massive deformation, respectively; and, the massless or massive CB model (\ref{eq127}), could be realized as a relevant deformation with the singlet operator $\mathcal{O}_2=(\mathcal{O}_1^{+})^2$, where with just keeping one boundary scalar, one should set the fermions and gauge fields to zero except the $U(1)$ part, for the latter two cases. As the same way, the massless or massive RF model (\ref{eq124}), could be realized as a relevant deformation with the singlet operator $\mathcal{O}_2^-$ in the ABJM original fermion action or a massive deformation of it, respectively; and, the massless or massive CF model (\ref{eq131}), could be realized as an irrelevant deformation with the singlet operator $\mathcal{O}_4=(\mathcal{O}_2^-)^{2}$, where with just keeping one boundary fermion, one should set the scalars and gauge fields to zero except the $U(1)$ part, for the latter two cases as well.

It is also good to mention that besides the operators composed of the fields of one type we have considered, one may also consider, depending on the cases at hand, correlated combinations of the bosons and fermions. \sloppy For instance, operators like \cite{Gaiotto-Yin2007} $\bar{\mathcal{O}}_2 \approx M_A^{\ B}\, \text{tr} \left(\psi^{A\,\dag} \psi_B + \frac{8 \pi}{k} Y^C Y_{[C} Y^A Y_{B]}^{\dag} \right)$ next to another (protected) chiral primary operator of $\bar{\mathcal{O}}_1 \approx M_A^{\ B}\, \text{tr}(Y^A Y_B^\dag)$, with $M_A^{\ B} = \text{diag}(1,1,-1,-1)$, are considered to study various (mass) deformations including \emph{Janus} solutions \cite{Kim2018} and SUSY Q-latices solutions \cite{Gauntlett2018} of the ABJM model; see also \cite{Arav2019} and \cite{Kim2019} for newer studies by the same authors, respectively. It is also notable that the fermion-bilinear deformation operators like $\tilde{\mathcal{O}}_2 \approx M_A^{\ B}\, \text{tr} \left(\psi^{A\,\dag} \psi_B \right)$, which are normally dual to the bulk pseudoscalars, and the scalar-bilinear deformation operators like $\bar{\mathcal{O}}_1$, which are normally dual to the bulk scalars, are also used to study \emph{boomerang RG flow} of the ABJM model in \cite{Donos2017}.

\section{Concluding Remarks}
From a consistent truncation of 11D SUGRA over $AdS_4 \times CP^3 \ltimes S^1/Z_k$ , including backreaction, we got equations for Higgs-like (pseudo)scalars of $m^2=0, -2, +4$. In particular, for the conformally coupled mode, we got a non-supersymmetric closed solution, and computed its contribution to the background action, which was in turn a nonzero finite value, confirming the instanton nature of the solution. After taking near the boundary behavior of the CC solution, we saw that with the mixed BC, it corresponded to a marginal triple-trace deformation of the dual boundary 3D field theory, and caused instability in that the associated effective potential was unbounded from below and indeed went to the infinite negative. On the other hand, we pointed out that the original scale- and parity- invariances were violated, and also all original SUSYs of the background SUGRA were broken because of the dynamics of the probe (anti)membranes. As a result, there were in general non-SUSY $SO(4)$-invariant instanton solutions, which could of course be realized in 3D $U(N)$ and $O(N)$ CS-matter theories; and to that end, we used the singlet sectors of the massless and mass deformed boundary models with keeping just an $U(1)$ (or $SO(2)$) part of the gauge group and one scalar or fermion depending on the particular model we considered.

More precisely, we saw that the boundary effective actions for the bulk mode with Dirichlet and Neumann BCs, were free or regular fermion and boson models, which with double-trace deformations of the corresponding $\Delta_{\pm}=2, 1$ operators went to critical fermion and boson models, respectively. After that, we wrote exact instanton solutions and computed the action corrections, from the EoMs derived from the corresponding actions of the models. In addition, we confirmed the state-operator AdS$_4$/CFT$_3$ correspondence for all the solutions and operators we had in our setups; Especially, we saw that a dimention-4 operator, which came from a double-trace deformation originated from the Neumann BC for the bulk CC (pseudo)scalar, corresponded to the massive bulk mode under the Dirichlet BC as well. Then, we confirmed the Bose-Fermi duality between RB and CF and also between RF and CB models at the level of the solutions. Next, we discussed briefly the similarities and relations of our solutions and setups to the Vasiliev HS and ABJM models as well as a few more related studies.

Then, as a complementary study, it will be interesting to find or build dual boundary effective actions corresponding to the main bulk scalar equation (\ref{eq06}), for various (pseudo)scalar modes, by standard holographic renormalization methods as, for instance, in \cite{deHaroSolodukhinSkenderis} and \cite{Bianchi3} and also \cite{Papadimitriou2006}.

As another point, although we remind that our solution here is a well-known example of a non-SUSY unstable AdS vacuum, agreed with the conjectures in \cite{OoguriVafa016} and \cite{Freivogel-Kleban2016}, it is good to explore more the stability of various possible configurations and solutions here as well. An original study on the stability analyzes, in these setups, is  \cite{Hertog-Horowitz2004a} where it is argued that a theory like ours is stable provided that the associated potential has a global minimum (namely there should be a bounded energy from below, which of course was not the case with our triple-trace deformation (\ref{eq119}) and so, we had instability) and admits a special superpotential; for other related studies, see \cite{Amsel2007} and \cite{Faulkner2010}. There is also a similar discussion in \cite{LucaVecchi}, where it is argued that for the $AdS_4$ stability, the effective potential or deformation must be $V_{eff.} \geq 0$; There, it is also argued that for a mode like ours, in the interval $-9/4 \leq m^2 \leq -5/4$ , the deformed theory flows between two fixed points of the RG with a resonance at the scale settling the transition between two CFT's; and that the latter resonance does not backreact on the geometry and is characteristic of the geometries interpolating between two $AdS_4$ spaces, in agreement with our discussions.

Finally, it will be interesting to work more on instanton solutions for the new 3D CS-matter models and BF dualities (for instance, those in \cite{Aharony2018} and \cite{Choudhury2018}) as well as HS theories and their duals (see, for instance, \cite{Sezgin-Sundell2005-01} and \cite{Giombi2019} for related studies).

\section{Acknowledgments}
I would like to thank the members of the High Energy, Cosmology and Astroparticle Physics (HECAP) section of the Abdus Salam International Centre for Theoretical Physics (ICTP), in particular A. Dabholkar and K. Papadodimas, for invitation, hospitality and scientific discussions during my recent visit there, where some parts of this study were fulfilled. I would also like to thank P. Creminelli, B. Acharya, E. Gava, F. Quevedo, G. Thompson, K. S. Narain, P. Putrov and the visitors A. Pankov, S. de Alwis, K. Narayan and M. M. Sheikh-Jabbari at the center and also M. Frasca and A. Imaanpur for related scientific discussions.


\begin{appendices}

\section{11D Supergravity Action and Equations} \label{Appendix.A1}
In this study, we use the Euclidean version of the bosonic part of the 11D SUGRA action as
\begin{equation}\label{eq100}
  S_{11}^E = -\frac{1}{2 \kappa_{11}^2} \left[ \int d^{11}x \, \sqrt{g} \, \mathcal{R} + \frac{1}{2} \int \left({G}_4 \wedge \ast_{11} {G}_4 - \frac{i}{3}\, {\mathcal{A}}_3 \wedge {G}_4 \wedge {G}_4 \right) \right],
\end{equation}
where, in general, $2 \kappa_{D}^2 = 18 \pi \mathcal{G}_{D}=\frac{1}{2\pi} (2\pi l_p)^9$, with $\kappa_{D}$, $\mathcal{G}_{D}$ and $l_p$ as the D-dimensional gravitational constant, Newton's constant and Plank length, respectively; and $G_4= d\mathcal{A}_3$.

The resultant EoMs for $\mathcal{A}_3$ and $g_{MN}$ read
\begin{equation}\label{eq101}
  d \ast_{11} G_4 -\frac{i}{2}\, G_4 \wedge G_4 \equiv d \hat{G}_7 =0,
\end{equation}
\begin{equation}\label{eq102}
    \mathcal{R}_{MN} - \frac{1}{2} g_{MN}\, \mathcal{R} = 8 \pi\, \mathcal{G}_{11}\, T_{MN}^{{G}_4},
\end{equation}
with $M, N,...$ for the 11D space-time indices and
\begin{equation}\label{eq102a}
 T_{MN}^{{G}_4} = \frac{1}{4!} \left[4\, {G}_{MPQR}\, {G}_N^{PQR} - \frac{1}{2} g_{MN}\, {G}_{PQRS}\, {G}^{PQRS} \right],
\end{equation}
respectively.

It is notable that the bosonic fields of 11D SUGRA include the graviton $g_{MN}$ with 44 degrees of freedom and the rank-3 antisymmetric tensor $\mathcal{A}_3$ with 84 degrees of freedom, so that the total 128 degrees of freedom equal to degrees of freedom of the gravitino $\Psi_M$, which in turn is the only necessary fermion field (a 32-component Majorana spinor) in the theory to preserve supersymmetry. Beside the extended objects, coupled to the gauge field $\mathcal{A}_3$ and its 11D dual, which are (electrically charged) M2-branes and (magnetically charged) M5-branes, there are also M9-branes and a pair of purely gravitational objects namely gravitational wave ($M \mathcal{W}$) and Kaluza-Klein ($\mathcal{K} \mathcal{K}$) monopoles. In addition, the electrical or page charge and magnetic or topological charge
\begin{equation}\label{eq103}
 Q_e = \frac{1}{\sqrt2\, \kappa_{11}^2} \int \hat{G}_7, \qquad Q_m=\frac{1}{\sqrt2\, \kappa_{11}^2} \int G_4,
\end{equation}
obeying the Dirac quantization condition $Q_e Q_m = 2 \pi Z$, come from the $\mathcal{A}_3$ equation (\ref{eq101}) and the Bianchi identity $dG_4=0$, respectively.

Further, having a classical solution, the killing spinors $\epsilon$ control the numbers of supersymmetries from zeroing the gravitino variation
\begin{equation}\label{eq104}
     \delta \Psi_M = D_M \epsilon -\frac{1}{128} \left(\Gamma_M^{PQRS} - 8\, \delta_M^P\, \Gamma^{QRS} \right)\, G_{PQRS}\, \epsilon =0,
\end{equation}
where $D_M$ is for the covariant derivative and $\Gamma^{M_1 M_2 ...}$  are the anti-symmetrized higher dimensional gamma matrices.

\section{\large{Matching with Setups and Results of \cite{Gauntlett03}}} \label{Appendix.A22}
As we mentioned at the end of subsection \ref{subsection02-01}, a general truncation of 11D SUGRA using the anstazs
\begin{equation}\label{eq122a}
ds_{11}^2 = ds_4^2 + e^{2 U} ds^2 ({K E_6}) + e^{2 V} \left( \eta + A_1 \right)^2,
\end{equation}
\begin{equation}\label{eq122b}
 \begin{split}
 \tilde{G}_4 = & f\, \texttt{vol}_4 + H_3 \wedge \left( \eta + A_1 \right) + H_2 \wedge J + H_1 \wedge J \wedge  \left( \eta + A_1 \right) + 2 h\, J \wedge J \\
   &  + \sqrt{3}\, \big[\chi_1 \wedge \Omega + \chi\, \left( \eta + A_1 \right) \wedge \Omega + c.c. \big]
 \end{split}
\end{equation}
is worked out in \cite{Gauntlett03}, where $ds_4^2$ and $ds^2 ({K E_6})$ (a K\"{a}hler-Einstein metric) are the full metrics of $AdS_4$ and $CP^3$ spaces respectively, $U, V, h, f$ are scalar fields, $A_1$ is an 1-form, $H_p\, (p=1,2,3)$ are $p$-forms, $\chi_1$ is a complex 1-form and $\chi$ is a complex scalar on the external 4D space; the $(3,0)$-form $\Omega$ defines the complex structure on $CP^3$ and "$c.c.$" denotes complex conjugate.

Given the above ansatzs, the details of KK reduction, after solving the 11D SUGRA equations, are given in Appendix B of \cite{Gauntlett03}. By comparing the results there and our results, we see that the solutions match with $U = V = \chi = A_1 = B_1 = B_2 = 0$ (noting that $H_1=dh,\, H_2=dB_1+2B_2 +hF_2,\, H_3=dB_2,\, F_2=dA_1,\, \chi_1=-\frac{i}{4} d\chi- A_1 \chi$), $\eta=2 e_7$, $\texttt{vol}_4 =\frac{R^4}{16}  \mathcal{E}_4$ and
\begin{equation}\label{eq123}
     f = 6\, f_1, \quad h= 4\, R^4 f_3, \quad dh= - R^4\, df_2,
\end{equation}
\begin{equation}\label{eq124}
     f = \frac{6}{R^7} \left(\epsilon + h^2 \right), \quad \epsilon =\pm\, C_3\, R^6,
\end{equation}
where the last equation comes from (B.12) of \cite{Gauntlett03}. In particular, from the equation (B.11) there, we read \begin{equation}\label{eq125}
     \ast_4 d (\ast_4 dh)- (16 + 24 \epsilon)\, h - 2 \times 12\, h^3 =0,
\end{equation}
which is indeed the equation for $f_3$ in (\ref{eq03}) up to some scaling and considering that $R=1$ is set; and also note that with $G_4 \rightarrow i G_4$, the Euclidean equation (\ref{eq101}) goes to that in \cite{Gauntlett03}. It is also noticeable that from the latter equation, we get the modes $m^2 R_{AdS}^2=-2$ with $\epsilon = -1$ (SW) and $m^2 R_{AdS}^2=10$ with $\epsilon = 1$ (WR), which were recently discussed in \cite{Me6} and \cite{Me7} respectively.\\
It is also noticeable that the 4D equation (\ref{eq125}) for $h$ can also be obtained from the string-frame action (2.10) of \cite{Gauntlett03}, which with our adjustments reads
\begin{equation}\label{eq126}
     S_{4E} = \int d^4x\, \sqrt{g_4}\, R^7 \left[- \left(\mathcal{R}_4-\mathcal{X} \Lambda \right) + \frac{3}{2 R^6} \left(\nabla h \right)^2 + \frac{3 h^2}{R^8} \left( 8+ \frac{12 \epsilon}{R^6}+ \frac{6 h^2}{R^6}\right) \right],
\end{equation}
where $\mathcal{R}_4$ is the scalar curvature of $EAdS_4$ and $\mathcal{X} \Lambda = \frac{1}{R^2} \left(-42+ 18 \epsilon^2 \right)$ with $\Lambda$ as the cosmological constant, noting that for $\epsilon= \pm 1$ there is the well-known result $\mathcal{X}=2$.

On the other hand, with the 11D Einstein equations
\begin{equation}\label{eq127}
 \mathcal{R}_{MN}=  \frac{1}{12} {G}_{MPQR}\, {G}_N^{PQR} - \frac{1}{144} g_{MN}\, {G}_{PQRS}\, {G}^{PQRS},
\end{equation}
for the components $\mathcal{R}_{\mu \nu}$, $\mathcal{R}_{mn}$ and $\mathcal{R}_{77}$, the equations (B.19), (B.21) and (B.22) of \cite{Gauntlett03} follow  respectively, which in turn correspond to the equations (\ref{eq08aa}), (\ref{eq08c}) and (\ref{eq08b}), when written for $f_3$ and the above adjustments are considered. Finally, as argued also in \cite{Gauntlett03}, becuase all dependencies on the internal 7D space is dropped out of the 11D equations and we are left with the equations for the 4D fields, therefore the ansatzs (\ref{eq01}) and (\ref{eq02}) defines a consistent Kaluza-Klein truncation.

\section{\large{To Compute the Instanton (\ref{eq44a}) Correction to the Background Action}} \label{Appendix.A2}
Here we compute the correction to the original action because of the bulk instanton solution (\ref{eq44a}). As the background geometry is unchanged, the right parts of the bosonic 11D SUGRA action (\ref{eq100}) for our purpose, are the second and third terms.

On the other hand, from the ansatz (\ref{eq01}), we get the 11D dual 7-form
\begin{equation}\label{eq47}
  {G}_7 = R^7\, {f}_1\, J^3 \wedge e_7 + R^5\, \ast_4 df_2 \wedge J^2 + R^7\, f_3\, \mathcal{E}_4 \wedge J \wedge e_7;
\end{equation}
and write
\begin{equation}\label{eq48}
 {G}_4 =  d{\mathcal{A}}_3, \quad {\mathcal{A}}_3 = \tilde{\mathcal{A}}_3^{(0)} + \left(8\, R^8 \right) \left(f_3\, J \wedge e_7 \right), \quad \tilde{G}_4^{(0)} = d\tilde{\mathcal{A}}_3^{(0)} = \frac{3}{8} R^4 f_1\, \mathcal{E}_4.
\end{equation}
With these at hand, using (\ref{eq03a}) and (\ref{eq03b}) and after some math manipulations, we arrive at
\begin{equation}\label{eq49}
  \tilde{S}_{11}^E = - \frac{R^{11}}{32\, \kappa_{11}^2} \int \bigg[-3\, c_3^2 + R^2\, (\partial_{\mu} f_2)(\partial^{\mu} f_2)+ 4\, f_2^2 + 12\, R^2\, f_2^4 \bigg]\, \mathcal{E}_4 \wedge J^3 \wedge e_7,
\end{equation}
where we have discarded the surface term of $ d\left(f_2^2\, \mathcal{A}_3^{0} \right)$ in the integrand that, as a total derivative, does not contribute to the equations; and that $\mu, \nu,...$ are $AdS_4$ space indices. It is also important to note that the mode $m^2 R_{AdS}^2=-2$, which we are interested in its associated solution (\ref{eq44a}) here, is also realizable in the SW background $C_2=0, C_3=1$ of (\ref{eq06a}) \footnote{It should be noted that for real $C_3> 1$ in the SW case, the same (CC) mode could also be realized with $C_2\neq 0$ in the main equation (\ref{eq06}); and that the action correction is computed in a similar way.} and so, we use the same setting for the equation (\ref{eq08g}) stemmed from combining the equations (\ref{eq06}) and (\ref{eq08c}).

To continue computing (\ref{eq49}), we use
\begin{equation}\label{eq49a}
  \texttt{vol}_4 = \frac{R^4}{16} \int \mathcal{E}_4, \quad  \mathcal{E}_4 = -\frac{du}{u^4} \wedge  dx \wedge dy \wedge dz,
\end{equation}
\begin{equation}\label{eq49b}
\texttt{vol}_7 = \frac{R^7}{3!} \int J^3 \wedge e_7= \frac{\pi^4\, R^7}{3\, k},
\end{equation}
where $\texttt{vol}_4$ and $\texttt{vol}_7$ are the full external and internal volumes, and
\begin{equation}\label{eq49c}
  {R}/{l_p}= \left(k\, N\, 2^5 \pi^2 \right)^{1/6}, \quad 4 \pi \kappa_{11}^2 = \left(2 \pi\, l_p \right)^9 \Rightarrow \kappa_{11}^2 = \frac{16}{3} \pi^5 \left(\frac{R^9}{3\, k^3} \right)^{1/2},
\end{equation}
where the first relation from the left is from \cite{ABJM}.

Then, we note that the first term on the RHS of (\ref{eq49}) is for the SW background realized with ${C}_3=1$, and to compute the remaining part, we use the instanton solution (\ref{eq44a}) (with $C_2=0$) with the 3D spherical coordinates.

\section{\large{Solutions For the Main Equation with} \textbf{$m^2=-2, +4$}} \label{Appendix.A3}
For the CC (pseudo)scalar $m^2=-2$, with the parameters $ C_3=\frac{13}{12}$, $C_2=\frac{1}{24\, \sqrt{2}}$, in the SW version of the main equation (\ref{eq06}), the resultant equation reads 
\begin{equation}\label{eq08ggaa}
    \Box_4\, f + 2\, f + 3\sqrt{2}\, f^2 - 24\, f^3 = 13/(72 \sqrt2). 
\end{equation}
To have an approximate solution to the equation, besides the one based on the closed solution in \cite{Me7}, we rewrite the homogeneous part of the equation (\ref{eq08ggaa}) (discarding the non-homogenous $F$ term, which of course adds just a non-dynamical or constant term to the final solution), making use of the conformal flatness of the Euclidean AdS space and the scaling of (\ref{eq42}), as
\begin{equation}\label{eq53hhh}
 \left(\partial_i \partial_i + \partial_u \partial_u \right)\, g_{i+1} = \sum_{i=0}^{\infty} A_i,
\end{equation}
with
\begin{equation}\label{eq53ii}
A_0 = -\frac{\delta}{u} g_0^2 + 24\, g_0^3 , \quad  A_1= -\frac{2 \delta}{u} g_0\, g_1 + 72\, g_0^2\, g_1, \ ... \, ,
\end{equation}
and the leading-order (LO) solution
\begin{equation}\label{eq53mmm}
\left[\frac{\partial^2}{\partial r^2} + \frac{2}{r} \frac{\partial}{\partial r} + \frac{\partial^2}{\partial u^2} \right] g_0(u,r)=0 \Rightarrow g_0(u,r) =\frac{{b}_0\, u}{\left[(u+a_0)^2 + r^2 \right]},
\end{equation}
where $r= \sqrt{x^2+y^2+z^2}$ with discarding the angular parts of the 3D spherical Laplacian for simplicity, and note that ${b}_0$ here is an arbitrary constant. Then, if we use the series expansion of the solution (\ref{eq53mmm}) about $u=0$ as
\begin{equation}\label{eq43aaa}
\bar{g}_0(u,r) =\frac{{b}_0}{(a_0^2+ r^2)} \left[1 - \frac{2\, a_0}{(a_0^2 + r^2)}\, u \right], \quad \bar{f}_0(u,r)= \bar{g}_0(u,r)\, u, 
\end{equation}
as the initial data in the recursion of the equation (\ref{eq53hhh}), we get the approximate solution
\begin{equation}\label{eq53jjj}
   f^{(2)}(u,r)= \bar{f}_0(u,r) - \frac{3\, {b}_0^2}{(a_0^2 + r^2)^2}\, \left[\frac{{b}_0}{a_0} -\sqrt{2}\, (1-\ln(u)) \right]\, u^2+ O(u^3),
\end{equation}
up to the second-order of the perturbation.

Similarly, another way to realize the {massive (pseudo)scalar $m^2=4$} is with $C_3=1, C_2=\frac{\sqrt{3}}{12}$, in the SW version of (\ref{eq06}) as
\begin{equation}\label{eq08ee}
    \Box_4\, f - 4\, f + 12\sqrt{3}\, f^2 - 24\, f^3 = 0,
\end{equation}
where $F=0$ is realized interestingly.

Then, to write a solution for the equation (\ref{eq08ee}), one may use the so-called Witten's solution \cite{Witten}, for its linear part in coordinate-space, as
\begin{equation}\label{eq50dddd}
     f_0(u,\vec{u})= \frac{8}{\pi^{2}} \left[\frac{u}{u^2+(\vec{u}-\vec{u}_0)^2} \right]^{4} \Rightarrow f_0(u\rightarrow 0,r)= \frac{8}{\pi^{2}} \left( \frac{u}{r^2} \right)^4 + O(u^6),
\end{equation}
where we have rewritten its behavior near the boundary on the right as well; and then, one can use perturbative methods to arrive at higher-order expansions around the LO solution.

On the other hand, we can use the \emph{self-similar reduction} method (see, for instance, \cite{Polyanin2012}), to solve the main NPDE (\ref{eq06}) through the scale-invariance of the variables 
\begin{equation}\label{eq41a}
   u\rightarrow S\, u, \quad r \rightarrow S^{s_1}\, r, \quad f\rightarrow  S^{s_2}\, f, \quad \Rightarrow  f(u,r)\rightarrow u^{s_3}\, \mathcal{F}(\xi), \quad \xi = {r}\,{u^{s_4}},
\end{equation}
where $S$ is the scaling-parameter, $s_1=-s_4=1$ and $s_2=s_3=0$ in this case. Doing so, the resultant NODE (for the SW version with $C_3\geq 1/3$) reads
\begin{equation}\label{eq41b}
 \left[ \left({\xi}^{2}+1 \right) {\frac{{d }^{2}}{{d}{\xi}^{2}}} + \frac{(2+ 4\, \xi^2)}{\xi}\,\frac{d}{d \xi}- m^2 \right] \mathcal{F}(\xi) + 6 \sqrt{3}\, m\, \mathcal{F}(\xi)^{2} - 24\, \mathcal{F}(\xi)^{3}=0;
\end{equation}
Then, a solution for its linear part becomes
\begin{equation}\label{eq41c}
 \mathcal{F}_0(\xi) = \frac{1}{\xi} \left[C_6\, P_{\Delta_+ - 2}(i \xi) + C_7\, Q_{\Delta_+ - 2} (i \xi) \right],
\end{equation} 
where $P_a(Z)$ and $Q_a(Z)$ are for the \emph{Legendre functions} of the first- and second- kind, respectively. Now, taking the LO solution $F_0$ in palace of the function in nonlinear terms of (\ref{eq41b}), we get a first-order solution in Legendre functions, again. Alternatively, one may also note that the equation (\ref{eq41b}) without the nonlinear terms is indeed the \emph{Riccati equation}, and that for special bulk modes, its solutions are in terms of {inverse trigonometric functions} \footnote{For instance, a solution with $m^2=4$ reads
\begin{equation}\label{eq41d}
 \mathcal{F}_0(\xi)= \left(\frac{1}{\xi}+ 3\, \xi \right) \left[1+ C_8\, \arctan(\xi) \right] + 3\, C_9.
\end{equation} }. 
Then, if we use the latter solutions, for a special bulk mode, to obtain $\mathcal{F}_1(\xi)$, the suitable terms (corresponding to the operators replying the bulk modes by the standard AdS/CFT dictionary) of the perturbative solution up to the first-order (noting $\mathcal{F}=\mathcal{F}_0 + \mathcal{F}_1 +... $), after substitution $\xi={r}/{u}$ and series expansion around $u=0$, is (\ref{eq41f}).

\section{\large{Basics of AdS$_4$/CFT$_3$ Dictionary For (pseudo)Scalars}} \label{Appendix.A4}
For a (pseudo)scalar in Euclidean $AdS$, near the boundary ($u=0$) behavior reads
\begin{equation}\label{eq50ee}
f(u\rightarrow 0, \vec{u}) \approx u^{\Delta_-} \left( \alpha(\vec{u})+ ...\right) +  u^{\Delta_+} \left(\beta(\vec{u}) + ...\right),
\end{equation}
where $2\, \Delta_{\mp}=3 \mp \nu$ with $\sqrt{9 + 4\, m^2}=2 \nu$ in $AdS_4$. Such a scalar could be quantized with the standard or Dirichlet ($\delta\alpha=0$), alternate or Neumann ($\delta\beta=0$) \cite{Witten}, \cite{Balasubramanian}, \cite{Dobrev1998}, \cite{KlebanovWitten} or Mixed boundary conditions \cite{Witten2}- see also \cite{Minces-Rivelles1999}- and that the latter BC could be considered as a multi-trace deformation of the field theory dual to the Neumann BC \cite{Papadimitriou02}. Indeed, for the (pseudo)scalars, the standard BC is used for any mass while regularity ($\Delta_+$ being real) and stability need satisfying the Breitenlohner–Freedman (BF) bound $m^2\geq m_{BF}^2=-9/4$ \cite{BreitenlohnerFreedman}; and at the same time, the alternate BC is always used for $-9/4 \leq m^2 \leq -5/4$ ensuring stability as well \cite{BreitenlohnerFreedman02}. We also remind that $\alpha$ and $\beta$ have holographic descriptions as source and vacuum expectation value (vev) of the one-point function of the operator $\Delta_+$ respectively, and conversely for the operator $\Delta_-$. Besides, with $m^2\geq -5/4$, only the $\beta$ mode is normalizable and so, dual operator with $\Delta_+$ and Dirichlet BC should be used; while for the masses in the range lower than the latter value and upper than the BF bound, both (now normalizable) modes and BCs are allowed.

For the standard and alternate boundary conditions, we use the following Euclidean AdS/CFT dictionary:
\begin{equation}\label{eq116}
   \begin{split}
   & \sigma \equiv \langle \mathcal{{O}}_{\Delta_+} \rangle_{\alpha} = - \frac{\delta W[\alpha]}{\delta\alpha} = \frac{1}{3} \beta, \quad \langle \mathcal{{O}}_{\Delta_-} \rangle_{\beta} = - \frac{\delta \tilde{W}[\sigma]}{\delta\sigma}= \alpha, \\
   & \ \ \ \ \ \ \ \ \ \ \ \ \ \ \ \tilde{W}[\sigma] = - W[\alpha] - \int d^3 \vec{u}\ \alpha(\vec{u})\, \sigma(\vec{u}), \\
   & \ \ W[\alpha] = -S_{on}[\alpha]=\Gamma_{eff.} [\alpha], \quad \tilde{W}[\sigma] = - \tilde{S}_{on}[\sigma]= \tilde{\Gamma}_{eff.} [\sigma],
   \end{split}
\end{equation}
where $S_{on}$ and $\tilde{S}_{on}$ are for the bulk $AdS_4$ on-shell actions, and $W[\alpha]$ and $\tilde{W}[\sigma]$ (Legendre transform of the other) are respectively generating functionals of the connected correlators of $\mathcal{{O}}_{\Delta_+}$ and $\mathcal{{O}}_{\Delta_-}$ of the boundary CFT$_3$, and also $\Gamma_{eff.} [\alpha]$ and $\tilde{\Gamma}_{eff.} [\sigma]$ (Legendre transform to each other) are respectively the effective actions of \emph{dual} CFT (with $\Delta_-$ quantization) and \emph{usual} CFT (with ${\Delta_+}$ quantization).

Still, one may consider deformations of the boundary CFT by a function ${V}(O_{\mp})$ of the local operators; see, for instance, \cite{Aharony2001}, \cite{Witten2}, \cite{Mueck2002}, \cite{Sever-Shomer2002}, \cite{Elitzur05}. In particular, for the mixed BC, also considered as a multi-trace deformation of the boundary theory dual to the Neumann BC \cite{Papadimitriou02}, we can write
\begin{equation}\label{eq117}
   \begin{split}
   & \ \ \ \ \ \ \ \ \ \ \ \ S_{on}^f[\alpha] = S_{on}[\alpha] + \int d^3 \vec{u}\ {V}(\alpha) \Rightarrow  \sigma_f=\sigma - \acute{{V}}(\alpha), \\
   & \ \ \ \ \ \ \ \ \ \ \ \ \ \ \ \ \ \ \ \ \ \ \ \ \Gamma_{eff.}^f [\alpha]= \Gamma_{eff.}[\alpha] +  \int d^3 \vec{u}\ {V}(\alpha),
   \end{split}
\end{equation}
where the $f$ index is for the associated quantity after the deformation, the prime on ${V}$ is for derivative wrt its argument $(\alpha)$, and in particular $\sigma_f$ is the new source for the operator $\mathcal{{O}}_-$ with the vev of $\alpha$. It is also mentionable that exchanging $\sigma\leftrightarrow\alpha$ in the above relations (\ref{eq117}) is valid for deformations with the Dirichlet BC.

\section{\large{Holographic Actions For the Bulk CC (pseudo)Scalar and Interpretations}} \label{Appendix.A5}
For a free minimally-coupled bulk (pseudo)scalar in the fixed or non-dynamical $AdS$ background, like that in (\ref{eq08g}), one can compute the boundary effective actions corresponding to the three boundary conditions and rules of the section \ref{Sec-06a} and particularly the Appendix \ref{Appendix.A4}; see \cite{Elitzur05} and  \cite{Papadimitriou02}. In particular, the regularized and renormalized effective actions corresponding to the Dirichlet and Neumann or mixed boundary conditions, using the Hamiltonian-Jacobi method and in two-derivative approximation, are computed in \cite{Papadimitriou02}, for the CC case with $\Delta_{\pm}=2,1$, to be
\begin{equation}\label{eq118a}
   \tilde{\Gamma}_{eff.} [\sigma] \approx \int d^3\vec{u}\ \left[ \frac{1}{2} \sigma^{-3/2}\, (\partial_i \sigma) (\partial^i \sigma) + \sigma^{1/2}\, \mathcal{R}_3  \right],
\end{equation}
\begin{equation}\label{eq118b}
 \begin{split}
   {\Gamma}_{eff.}^f [\alpha] = \int d^3\vec{u}\ & \left[ \frac{\alpha^{-1}}{8} (\partial_i \alpha) (\partial^i \alpha) + \frac{\alpha}{16} \mathcal{R}_3 + \tilde{c}_0\, \alpha^3 + V(\alpha) \right],
 \end{split}
\end{equation}
respectively, where $\mathcal{R}_3=6/R_0^2$ is the scalar curvature of the three-sphere $S^3$ (with $R_0\rightarrow \infty$ for the flat $R^3$ space or $S^3$ at infinity: $S^3_\infty$).

Then, we first note to the solution (\ref{eq44a}) for the case at hand that, with (\ref{eq46a}) and the rules (\ref{eq116}) and (\ref{eq117}), indeed corresponds to a triple-trace deformation of the boundary theory with
\begin{equation}\label{eq119}
   V(\alpha) = -\frac{1}{3} \hat{h}\, \alpha^3, \quad \hat{h}= \frac{2 a_0}{b_0};
\end{equation}
\footnote{Note that in \cite{HertogMaeda01}, \cite{Hertog-Horowitz2004} and \cite{Hertog-Horowitz2005}, a similar deformation is considered as $\beta= -\hat{h}\, \alpha^2$ and so, according to \cite{Witten2} or the rules of (\ref{eq116}), that formalism coincides with us as follows:
\begin{equation}\label{eq120}
   W[\alpha]= \frac{\hat{h}}{3} \int d^3 \vec{u}\ \alpha(\vec{u})^3, \quad  \tilde{W}[\sigma] = \frac{2 \check{h}}{3} \int d^3 \vec{u}\ \beta(\vec{u})^{3/2},
\end{equation}
where $\hat{h}=1/\check{h}^2$, and $\alpha$ and $\beta$ are vev of the operators $\mathcal{O}_{1}$ and $\mathcal{O}_{2}$, respectively.}. Therefore, from the last two terms of (\ref{eq118b}) and from (\ref{eq119}), we can write the holographic effective potential
\begin{equation}\label{eq121}
   V_{eff.}(\alpha) =\frac{1}{3} \left(\hat{h}_0 - \hat{h} \right) \alpha^3,
\end{equation}
where $\hat{h}_0=3 \tilde{c}_0$; and it is argued in \cite{Papadimitriou02} that for the 4D $\mathcal{N}=8$ gauged supergravity, $\tilde{c}_0=0$, where the bulk solution describes a sector of the Coulomb branch of the dual boundary theory. \footnote{See also \cite{Papadimitriou2006}, where the standard procedures of holographic renormalization, Hamiltonian-Jacobi and in particular \emph{fake superpotentials} are employed to get the boundary effective action replying to the classically marginal operator $\sim \mathcal{O}_1^3$, in large-$N$ limit.} \footnote{Look also at \cite{Mansi2009}, where for the CC (pseudo)scalar with the so-called $\phi^4$ self-interaction in $EAdS_4$ space (corresponding to the SW version of (\ref{eq03}) with $C_3=1$-also in \cite{Me8}- with $\lambda_4=192$), employing the \emph{stochastic quantization}, a similar boundary effective action is derived. In particular, it is interesting that the bulk coupling $\lambda_4=192$ stands for the boundary one, which is of course the UV fixed-point $g_6^*=192$ of the 3D tri-critical ($\varphi^6$) $O(N)$ vector model \cite{Pisarski1982}.}

In particular, it is shown in \cite{Elitzur05} that for $\hat{h}> \hat{h}_0$, there is a runway solution where the vev of the operator $\mathcal{O}_1$ (=$\alpha$) diverges in finite time (indeed, the rolling down of the scalar from the negative potential and producing a big-crunch at finite time) and the associated CFT$_3$ needs an UV-completion. In other words, to define a field theory, one must introduce an UV cut-off $\Lambda$ as the boundary relevant operator behaves badly at UV; and that (for $g_6 > g_6^c$; see below) the UV cut-off $\Lambda$ is not removed and an UV-completion is required to have a cut-off independent theory; and in this case, there is a big-crunch singularity in the bulk. In simpler language, the deformation or negative potential (\ref{eq121}) is unbounded from below and causes instability. \footnote{For situations with a mixed boundary condition like ours, it is argued in \cite{Hertog-Horowitz2004a} that the theory is stable provided that the associated potential has a global minimum.}

In fact, this setup was originally considered about 3D $O(N)$ vector models; see also \cite{Asnin2009} and \cite{CrapsHertog}. In this way, we may take $g_6^c=2\, \hat{h}_0$ and $g_6=2\, \hat{h}$, and note that in the classical approach, the sign of $g_6$ determines the system behavior and for the quantum case, being larger or smaller than $g_6^c=(4 \pi)^2$ \cite{Bardeen1984}; and that for $g_6 < g_6^c$, the tri-critical model is scale-invariant in LO, but above the critical value $g_6^c$, a nonzero mass parameter appears (i.e. $m_b^2\, \langle y \bar{y}\rangle$, with $y$ as the singlet boundary scalar) \footnote{Concerning the SI breaking, realized through either including mass terms or quantum corrections to the actions, we note that a massive deformation through a term like $m_b^2\, \texttt{tr}(Y^A Y_A^\dag)$, which is in turn a non-protected Konishi-like operator in the large-$N$ limit, may be turned on- see \cite{Gomiz} for the first example of such a deformation in ABJM model- we discussed the same deformation in \cite{Me8} that is valid here as well.} and so, the SI is broken because of $1/N$ corrections- For further related discussions, see \cite{Amit1985} and \cite{BardeenMoshe014}, where the SI breaking in 3D $U(N)$ plus CS theory is discussed, and also \cite{Omid2016}. On the other hand, in the Large-$N$ limit, non-perturbative effects destabilize the model for $g_6 > g_6^c$ and the boundary configurations decay in a finite time because of an infinite family of instantons on $S^3$ of the radius $R_0$- In other words, there is tunneling from the local minimum at $\alpha=0$ to the instability region at $\alpha\rightarrow \infty$. It should also be mentioned that for $g_6 > g_6^c$, only the neutral bounded states under $U(1)$, which in turn belong to neutral irreducible reps of $SU(N)$, are remained  \cite{RabinoviciSmolkin011}.

\end{appendices}




\begin{thebibliography}{99}
\bibitem{Me3} M. Naghdi, \textit{"New Instantons in AdS$_4$/CFT$_3$ from D4-Branes Wrapping Some of CP$^3$"}, Phys. Rev. D 88, 026013 (2013),\href{http://arxiv.org/abs/1302.5294}{[arXiv:1302.5294 [hep-th]]}.

\bibitem{Me4} M. Naghdi, \textit{"Marginal Fluctuations as Instantons on M2/D2-Branes"}, Eur. Phys. J. C 74, 2826 (2014),
    \href{http://arxiv.org/abs/1302.5534}{[arXiv:1302.5534 [hep-th]]}.

\bibitem{Me5} M. Naghdi, \textit{"Dual localized objects from M-branes over $AdS_4 \times S^7/Z_k$"},  Class. Quant. Grav. 32, 215018 (2015), \href{http://arxiv.org/abs/1502.03281}{[arXiv:1502.03281 [hep-th]]}.

\bibitem{Me6} M. Naghdi, \textit{"Non-Minimally Coupled Pseudoscalars in $AdS_4$ for Instantons in CFT$_3$"}, Class. Quant. Grav. 33, 115005 (2016), \href{http://arxiv.org/abs/1505.00179}{[arXiv:1505.00179 [hep-th]]}.

\bibitem{ABJM} O. Aharony, O. Bergman, D. L. Jafferis and J. Maldacena, \textit{"$\mathcal{N}$=6 superconformal Chern-Simons matter theories, M2-branes and their gravity duals"}, JHEP 0810, 091 (2008), \href{http://arxiv.org/abs/0806.1218}{[arXiv:0806.1218 [hep-th]]}.

\bibitem{Me7} M. Naghdi, \textit{"Massive (pesudo)Scalars in AdS$_4$, SO(4) Invariant Solutions and Holography"}, Eur. Phys. J. Plus 133, 307 (2018), \href{http://arxiv.org/abs/1703.02765}{[arXiv:1703.02765 [hep-th]]}.

\bibitem{Me8} M. Naghdi, \textit{"A Truncation of 11-Dimensional Supergravity for Fubini-Like Instantons in AdS$_4/$CFT$_3$”}, Fortschr. Phys. 67, 1800044 (2018), \href{http://arxiv.org/abs/1708.02530}{[arXiv:1708.02530 [hep-th]]}.

\bibitem{DuffNilssonPope84} M. J. Duff, B. E. W. Nilsson and C. N. Pope, \textit{"Superunification from eleven dimensions"}, \href{http://www.sciencedirect.com/science/article/pii/0550321384905777?np=y} {Nucl. Phys. B 233, 433 (1984)}.

\bibitem{CrapsHertog} B. Craps, T. Hertog and N. Turok, \textit{"A multitrace deformation of ABJM theory"}, Phys. Rev. D 80, 086007 (2009), \href{http://arxiv.org/abs/0905.0709}{[arXiv:0905.0709 [hep-th]]}.

\bibitem{Vasiliev01} M. A. Vasiliev, \textit{"Higher spin gauge theories: Star product and AdS space"}, In "the many faces of the superworld: pp. 533-610", \href{http://arxiv.org/abs/hep-th/9910096}{[arXiv:hep-th/9910096]}.

\bibitem{Giombi2016} S. Giombi, I. R. Klebanov and Z. M.Tan, \textit{"The ABC of Higher-Spin AdS/CFT"}, Universe 4, 18 (2018), \href{http://arxiv.org/abs/1608.07611}{[arXiv:1608.07611 [hep-th]]}.

\bibitem{Klebanov-Polyakov2002} I. R. Klebanov and A. M. Polyakov, \textit{"AdS dual of the critical O(N) vector model"}, Phys. Lett. B 550, 213 (2002), \href{http://arxiv.org/abs/hep-th/0210114}{[arXiv:hep-th/0210114]}.

\bibitem{Sezgin-Sundell2002} E. Sezgin and P. Sundell, \textit{"Massless higher spins and holography"}, Nucl. Phys. B 644, 303 (2002), \href{http://arxiv.org/abs/hep-th/0205131}{[arXiv:hep-th/0205131]}, \href{https://www.sciencedirect.com/science/article/pii/S0550321303002670} {Erratum: Nucl. Phys. B 660, 403 (2003)}.

\bibitem{Chang2013} Chi-M. Chang, Sh. Minwalla, T. Sharma and Xi Yin, \textit{"ABJ triality: from higher spin fields to strings"}, J. Phys. A 46, 214009 (2013), \href{http://arxiv.org/abs/1207.4485}{[arXiv:1207.4485 [hep-th]]}.

\bibitem{Aharony2018} O. Aharony, S. Jain and Sh. Minwalla, \textit{"Bose-Fermi Chern-Simons dualities in the Higgsed phase"}, JHEP 1811, 177 (2018), \href{http://arxiv.org/abs/1804.08635}{[arXiv:1804.08635[hep-th]]}.

\bibitem{Choudhury2018} S. Choudhury, A. Dey, I. Halder, S. Jain, L. Janagal, Sh. Minwalla and N. Prabhakar, \textit{"Flows, fixed points and duality in Chern-Simons-matter theories"}, JHEP 1812, 058 (2018), \href{http://arxiv.org/abs/1808.03317}{[arXiv:1808.03317 [hep-th]]}.

\bibitem{FreundRubin1980} P. G. O. Freund and M. A. Rubin, \textit{"Dynamics of dimensional reduction"}, \href{https://www.sciencedirect.com/science/article/abs/pii/0370269380905900} {Phys. Lett. B 97, 233 (1980)}.

 \bibitem{Englert1982} F. Englert, \textit{"Spontaneous compactification of eleven-dimensional supergravity"}, \href{https://www.sciencedirect.com/science/article/abs/pii/0370269382906840} {Phys. Lett. B 119, 339 (1982)}.

 \bibitem{Pope1985} C. N. Pope, \textit{"An $SU(4)$ invariant compactification of $d = 11$ supergravity on a stretched seven-sphere"}, \href{https://www.sciencedirect.com/science/article/abs/pii/037026938590992X} {Phys. Lett. B 150, 352 (1985)}.

\bibitem{Gauntlett03} J. P. Gauntlett, S. Kim, O. Varela and D. Waldram, \textit{"Consistent supersymmetric Kaluza--Klein
    truncations with massive modes"}, JHEP 0904, 102 (2009), \href{http://arxiv.org/abs/0901.0676}{[arXiv:0901.0676 [hep-th]]}.

\bibitem{Duff99} M. J. Duff and J. T. Liu, \textit{"Anti-de Sitter black holes in gauged $\mathcal{N}=8$ supergravity"}, Nucl. Phys. B 554, 273 (1999), \href{http://arxiv.org/abs/hep-th/9901149}{[arXiv:hep-th/9901149]}.

\bibitem{Hrycyna2017} O. Hrycyna, \textit{"What $\xi$? Cosmological constraints on the non-minimal coupling constant"}, Phys. Lett. B 768, 218 (2017), \href{http://arxiv.org/abs/1511.08736}{[arXiv:1511.08736 [astro-ph.CO]]}.

\bibitem{Witten} E. Witten, \textit{"Anti-de Sitter space and holography"}, Adv. Theor. Math. Phys. 2, 253 (1998), \href{http://arxiv.org/abs/hep-th/9802150}{[arXiv:hep-th/9802150]}.

\bibitem{Polyanin2012} A. D. Polyanin and  V. F. Zaitsev, \textit{"Handbook of Nonlinear Partial Differential Equations"}, \href{https://www.taylorfrancis.com/books/9780429150371} {Chapman and Hall/CRC Press, 2nd Edition (2012)}.

\bibitem{Balasubramanian} V. Balasubramanian, P. Kraus and A. Lawrence, \textit{"Bulk vs boundary dynamics in anti-de Sitter spacetime"}, Phys. Rev. D 59, 046003 (1999), \href{http://arxiv.org/abs/hep-th/9805171}{[arXiv:hep-th/9805171]}.

\bibitem{Dobrev1998} V. K. Dobrev, \textit{"Intertwining operator realization of the AdS/CFT correspondence"}, Nucl. Phys. B 553, 559 (1999), \href{http://arxiv.org/abs/hep-th/9812194}{[arXiv:hep-th/9812194]}.

\bibitem{KlebanovWitten} I. R. Klebanov and E. Witten, \textit{"AdS/CFT correspondence and symmetry breaking"}, Nucl. Phys. B 556, 89 (1999), \href{http://arxiv.org/abs/hep-th/9905104}{[arXiv:hep-th/9905104]}.

\bibitem{Witten2} E. Witten, \textit{"Multi-trace operators, boundary conditions, and AdS/CFT correspondence"}, \href{http://arxiv.org/abs/hep-th/0112258}{[arXiv:hep-th/0112258]}.

\bibitem{Minces-Rivelles1999} P. Minces and V. O. Rivelles, \textit{"Scalar field theory in the AdS/CFT correspondence revisited"}, Nucl. Phys. B 572, 651 (2000), \href{http://arxiv.org/abs/hep-th/9907079}{[arXiv:hep-th/9907079]}.

\bibitem{Papadimitriou02} I. Papadimitriou, \textit{"Multi-trace deformations in AdS/CFT: Exploring the vacuum structure of the deformed CFT"}, JHEP 0705, 075 (2007), \href{http://arxiv.org/abs/hep-th/0703152}{[arXiv:hep-th/0703152]}.

\bibitem{BreitenlohnerFreedman} P. Breitenlohner and D. Z. Freedman, \textit{"Positive energy in anti-de Sitter backgrounds and gauged extended supergravity"}, \href{http://www.sciencedirect.com/science/article/pii/0370269382906438} {Phys. Lett. B 115, 197 (1982)}.

\bibitem{BreitenlohnerFreedman02} P. Breitenlohner and D. Z. Freedman, \textit{"Stability in gauged extended supergravity"},
    \href{http://www.sciencedirect.com/science/article/pii/0003491682901166}{Annals Phys. 144, 249 (1982)}.

\bibitem{Aharony2001} O. Aharony, M. Berkooz and E. Silverstein, \textit{"Multiple-trace operators and non-local string theories"}, JHEP 0108, 006 (2001), \href{http://arxiv.org/abs/hep-th/0105309}{[arXiv:hep-th/0105309]}.

\bibitem{Mueck2002} W. Mueck, \textit{"An improved correspondence formula for AdS/CFT with multi-trace operators"}, Phys. Lett. B 531, 301 (2002), \href{http://arxiv.org/abs/hep-th/0201100}{[arXiv:hep-th/0201100]}.

\bibitem{Sever-Shomer2002} A. Sever and A. Shomer, \textit{"A note on multi-trace deformations and AdS/CFT"}, JHEP 0207, 027(2002), \href{http://arxiv.org/abs/hep-th/0203168}{[arXiv:hep-th/0203168]}.

\bibitem{Elitzur05} S. Elitzur, A. Giveon, M. Porrati and E. Rabinovici, \textit{"Multitrace deformations of vector and adjoint theories and their holographic duals"}, JHEP 0602, 006 (2006), \href{http://arxiv.org/abs/hep-th/0511061}{[arXiv:hep-th/0511061]}.

\bibitem{Duff1985-2} M. J. Duff and C. N. Pope, \textit{"Consistent truncations in Kaluza-Klein theories"}, \href{https://www.sciencedirect.com/science/article/pii/0550321385901403?via%3Dihub}{Nucl. Phys. B 255, 355 (1985)}.

\bibitem{Duff1984-3} M. J. Duff, B. E. W. Nilsson and C. N. Pope, \textit{"The criterion for vacuum stability in Kaluza-Klein supergravity"}, \href{https://www.sciencedirect.com/science/article/abs/pii/0370269384912346} {Phys. Lett. B 139, 154 (1984)}.

\bibitem{Bena} I. Bena, \textit{"The M theory dual of a three-dimensional theory with reduced supersymmetry"}, Phys. Rev. D 62, 126006 (2000), \href{http://arxiv.org/abs/hep-th/000414}{[arXiv:hep-th/000414]}.

\bibitem{Avdeev1993} L. V. Avdeev, D. I. Kazakov abd I . N. Kondrashuk, \textit{"Renormalizations in supersymmetric and nonsupersymmetric non-abelian Chern—Simons field theories with matter"}, \href{https://www.sciencedirect.com/science/article/pii/055032139390151E}{Nucl. Phys. B 391, 333 (1993)}.

\bibitem{Aharony2012a} O. Aharony, G. G.-Ari and R. Yacoby, \textit{"Correlation functions of large $N$ Chern-Simons-Matter theories and bosonization in three dimensions"}, JHEP 1212, 028 (2012), \href{http://arxiv.org/abs/1207.4593}{[arXiv:1207.4593 [hep-th]]}.

\bibitem{Maldacena-Zhiboedov2012} J. Maldacena and A. Zhiboedov, \textit{"Constraining conformal field theories with a slightly broken higher spin symmetry"}, Class. Quant. Grav. 30, 104003 (2013), \href{http://arxiv.org/abs/1204.3882}{[arXiv:1204.3882 [hep-th]]}.

\bibitem{HertogMaeda01} T. Hertog and K. Maeda, \textit{"Black holes with scalar hair and asymptotics in N=8 supergravity"},
    JHEP 0407, 051 (2004), \href{http://arxiv.org/abs/hep-th/0404261}{[arXiv:hep-th/0404261]}.

\bibitem{Hertog-Horowitz2004} T. Hertog and G. T. Horowitz, \textit{"Towards a big crunch dual"}, JHEP 0407, 073 (2004), \href{http://arxiv.org/abs/hep-th/0406134}{[arXiv:hep-th/0406134]}.

\bibitem{Hertog-Horowitz2005} T. Hertog and G. T. Horowitz, \textit{"Holographic description of AdS cosmologies"}, JHEP 04, 005 (2005), \href{http://arxiv.org/abs/hep-th/0503071}{[arXiv:hep-th/0503071]}.

\bibitem{Papadimitriou2006} I. Papadimitriou, \textit{"Non-supersymmetric membrane flows from fake supergravity and multi-trace deformations"}, JHEP 0702, 008 (2007), \href{http://arxiv.org/abs/hep-th/0606038}{[arXiv:hep-th/0606038]}.

\bibitem{Mansi2009} D. S. Mansi, A. Mauri and A. C. Petkou, \textit{"Stochastic quantization and AdS/CFT"}, Phys. Lett. B 685, 215 (2010), \href{http://arxiv.org/abs/0912.2105}{[arXiv:0912.2105 [hep-th]]}.

\bibitem{Pisarski1982} R. D. Pisarski, \textit{"Fixed-point structure of $(\varphi^6)_3$ at large
$N$"}, \href{https://journals.aps.org/prl/abstract/10.1103/PhysRevLett.48.574}{Phys. Rev. Lett. 48, 574 (1982)}.

\bibitem{Hertog-Horowitz2004a} T. Hertog and G. T. Horowitz, \textit{"Designer gravity and field theory effective potentials"}, Phys. Rev. Lett. 94, 221301 (2005), \href{http://arxiv.org/abs/hep-th/0412169}{[arXiv:hep-th/0412169]}.

\bibitem{Asnin2009} V. Asnin, E. Rabinovici and M. Smolkin, \textit{"On rolling, tunneling and decaying in some large N vector models"}, JHEP 0908, 001 (2009), \href{http://arxiv.org/abs/0905.3526}{[arXiv:0905.3526 [hep-th]]}.

\bibitem{Bardeen1984} W. A. Bardeen, M. Moshe and M. Bander, \textit{"Spontaneous breaking of scale invariance and the ultraviolet fixed point in $O(N)$-Symmetric ($\varphi_3^6$) Theory"}, \href{http://journals.aps.org/prl/abstract/10.1103/PhysRevLett.52.1188}{Phys. Rev. Lett. 52, 1188 (1984)}.

\bibitem{Gomiz} J. Gomis, D. Rodriguez-Gomez, M. Van Raamsdonk and H. Verlinde, \textit{"A massive study of M2-brane proposals"}, JHEP 0809, 113 (2008), \href{http://arxiv.org/abs/0807.1074}{[arXiv:0807.1074 [hep-th]]}.

\bibitem{Amit1985} D. J. Amit and E. Rabinovici, \textit{"Breaking of scale invariance in $\varphi^6$ theory: Tricriticality and critical end-points"}, \href{https://www.sciencedirect.com/science/article/pii/0550321385903517?via%3Dihub} {Nucl. Phys. B 257, 371 (1985)}.

\bibitem{BardeenMoshe014} W. A. Bardeen and M. Moshe, \textit{"Spontaneous breaking of scale invariance in a $D=3\ U(N)$ model with Chern-Simons gauge field"}, JHEP 1406, 113 (2014), \href{http://arxiv.org/abs/1402.4196}{[arXiv:1402.4196 [hep-th]]}.

\bibitem{Omid2016} H. Omid, G. W. Semenoff, and L. C. R. Wijewardhana, \textit{"Light dilaton in the large $N$ tricritical $O(N)$ model"}, Phys. Rev. D 94, 125017 (2016), \href{http://arxiv.org/abs/1605.00750}{[arXiv:1605.00750[hep-th]]}.

\bibitem{RabinoviciSmolkin011} E. Rabinovici and M. Smolkin, \textit{"On the dynamical generation of the Maxwell term and scale invariance"}, JHEP 1107, 040 (2011), \href{http://arxiv.org/abs/1102.5035}{[arXiv:1102.5035 [hep-th]]}.

\bibitem{Fubini1} S. Fubini, \textit{"A new approach to conformal invariant field theories"},
     \href{http://link.springer.com/article/10.1007%2FBF02785664}{Nuovo Cim. A 34, 521 (1976)}.

\bibitem{Loran2} F. Loran, \textit{"Fubini vacua as a classical de Sitter vacua"}, Mod. Phys. Lett. A 22, 2217 (2007), \href{http://arxiv.org/abs/hep-th/0612089}{[arXiv:hep-th/0612089]}.

\bibitem{deharo06} S. de Haro and A. C. Petkou, \textit{"Instantons and conformal holography"}, JHEP 0612, 076 (2006), \href{http://arxiv.org/abs/hep-th/0606276}{[arXiv:hep-th/0606276]}.

\bibitem{BarbonRabinovici011} J. L. F. Barbon and E. Rabinovici, \textit{"AdS crunches, CFT falls and cosmological complementarity"}, JHEP 1104, 044 (2011), \href{http://arxiv.org/abs/1102.3015}{[arXiv:1102.3015 [hep-th]]}.

\bibitem{SmolkinTurok} M. Smolkin and N. Turok, \textit{"Dual description of a 4d cosmology"},
    \href{http://arxiv.org/abs/1211.1322}{[arXiv:1211.1322 [hep-th]]}.

\bibitem{Coleman1980a} S. R. Coleman and  F. De Luccia, \textit{"Gravitational effects on and of vacuum decay"}, \href{https://journals.aps.org/prd/abstract/10.1103/PhysRevD.21.3305} {Phys. Rev. D 21, 3305 (1980)}.

\bibitem{Witten3} E. Witten, \textit{"Instability of the Kaluza-Klein vacuum"}, \href{http://www.sciencedirect.com/science/article/pii/0550321382900074}{Nucl. Phys. B 195, 481 (1982)}.

\bibitem{Brown-Dahlen2011} A. R. Brown and A. Dahlen, \textit{"On “nothing” as an infinitely negatively curved spacetime"}, Phys. Rev. D 85, 104026 (2012), \href{http://arxiv.org/abs/1111.0301}{[arXiv:1111.0301 [hep-th]]}.

\bibitem{Barbon-Rabinovici2011} J. L. F. Barbon and E. Rabinovici, \textit{"Holography of AdS vacuum bubbles"}, \href{https://www.sciencedirect.com/science/article/pii/S0920563211004282?via%3Dihub} {Nucl. Phys. Proc. Suppl. 216, 121 (2011)}.

\bibitem{Harlow2010} D. Harlow, \textit{"Metastability in anti de Sitter space"},
    \href{http://arxiv.org/abs/1003.5909}{[arXiv:1003.5909 [hep-th]]}.

\bibitem{Barbon-Rabinovici2013} J. L. F. Barbon and E. Rabinovici, \textit{"Conformal complementarity maps"}, JHEP 1312, 023 (2013), \href{http://arxiv.org/abs/1308.1921}{[arXiv:1308.1921 [hep-th]]}.

\bibitem{Kumar-Vaganov2015} S. P. Kumar and V. Vaganov, \textit{"Probing crunching AdS cosmologies"}, JHEP 1602, 026 (2016), \href{http://arxiv.org/abs/1510.03281}{[arXiv:1510.03281 [hep-th]]}.

\bibitem{BzowskiHertog} A. Bzowski, T. Hertog and M. Schillo, \textit{"Cosmological singularities encoded in IR boundary correlations"}, JHEP 1605, 168 (2016), \href{http://arxiv.org/abs/1512.05761}{[arXiv:1512.05761 [hep-th]]}.

\bibitem{Kumar-Vaganov2018} S. P. Kumar and V. Vaganov, \textit{"Nonequilibrium dynamics of the $O(N)$ model on $dS_3$ and AdS crunches"}, JHEP 1803, 092 (2018), \href{http://arxiv.org/abs/1802.08202}{[arXiv:1802.08202 [hep-th]]}.

\bibitem{Maldacena010} J. Maldacena, \textit{"Vacuum decay into Anti de Sitter space"}, \href{http://arxiv.org/abs/1012.0274}{[arXiv:1012.0274 [hep-th]]}.

\bibitem{Akdeniz1979} K. G. Akdeniz and A. Smailagi\'{c}, \textit{"Classical solutions for fermionic models"},
     \href{http://link.springer.com/article/10.1007/BF02776595}{Nuovo Cim. A 51, 345 (1979)}.

\bibitem{Wilson-Fisher1972} K. G. Wilson and M. E. Fisher, \textit{"Critical exponents in 3.99 dimensions"}, \href{https://journals.aps.org/prl/abstract/10.1103/PhysRevLett.28.240}{Phys. Rev. Lett. 28, 240 (1972)}.

\bibitem{Giombi2017} S. Giombi, V. Gurucharan, V. Kirilin, S. Prakash and E. Skvortsov, \textit{"On the higher-spin spectrum in large $N$ Chern-Simons vector models"}, JHEP 1701, 058 (2017), \href{http://arxiv.org/abs/1610.08472}{[arXiv:1610.08472 [hep-th]]}.

\bibitem{Lipatov1977} L. N. Lipatov, \textit{"Divergence of the perturbation-theory series and the quasi-
classical theory"}, \href{http://inspirehep.net/record/111423?ln=en} {Sov. Phys. JETP 45, 216 (1977); Zh. Eksp. Teor. Fiz. 72 , 411 (1977)}.

\bibitem{brzein1977} E. Brézin, J. C. Le Guillou and J. Zinn-Justin, \textit{"Perturbation theory at large order. I. The $\phi^{2N}$ interaction"}, \href{https://journals.aps.org/prd/abstract/10.1103/PhysRevD.15.1544} {Phys. Rev. D 15, 1544 (1977)}.
\bibitem{McKane-Wallace1978} A. J. McKane and D. J. Wallace, \textit{"Instanton calculations using dimensional regularisation"}, \href{https://iopscience.iop.org/article/10.1088/0305-4470/11/11/013} {J. Phys. A 11, 2285 (1978)}.

\bibitem{Actor1982} A. Actor, \textit{"Nonexistence of multiple-instanton solution of the massless 4D $\phi^4$ theory"}, \href{https://onlinelibrary.wiley.com/doi/abs/10.1002/prop.19820300803} {Fortschr. Phys. 30, 437 (1982)}.

\bibitem{Justin1982} J. Zinn-Justin, \textit{"The principles of instanton calculus: A few applications"}, \href{http://inspirehep.net/record/182566?ln=en}{In Recent Advances in Field Theory, Les Houches, Session XXXIX, edited by J.-B. Zuber and R. Stora (North Holland, Amsterdam), pp. 39-172 (1982)}.

\bibitem{Coleman1978a} S. Coleman, V. Glaser and A. Martin, \textit{"Action minima among solutions to a class of Euclidean scalar field equations"}, \href{https://link.springer.com/article/10.1007%2FBF01609421}{Commun. Math. Phys. 58, 211 (1978)}.

\bibitem{Affleck1} I. Affleck, \textit{"On constrained instantons"},
    \href{http://www.sciencedirect.com/science/article/pii/0550321381903072}{Nucl. Phys. B 191, 429 (1981)}.

\bibitem{Nielsen1999} M. Nielsen and N. K. Nielsen, \textit{"Explicit construction of constrained instantons"}, Phys. Rev. D 61, 105020 (2000), \href{http://arxiv.org/abs/hep-th/9912006}{[arXiv:hep-th/9912006]}.

\bibitem{Belavin} A. A. Belavin, A. M. Polyakov, A. S. Shvarts and Yu. S. Tyupkin, \textit{"Pseudoparticle solutions of the Yang-Mills equations"}, \href{http://www.sciencedirect.com/science/article/pii/037026937590163X}{Phys. Lett. B 59, 85 (1975)}.

\bibitem{Hooft1976} G. 't Hooft, \textit{"Computation of the quantum effects due to a four-dimensional pseudoparticle"}, \href{https://journals.aps.org/prd/abstract/10.1103/PhysRevD.14.3432} {Phys. Rev. D 14, 3432 (1976)}, \href{https://journals.aps.org/prd/abstract/10.1103/PhysRevD.18.2199.3} {Erratum: Phys. Rev. D 18, 2199 (1978)}.

\bibitem{Alfaro1976} V. de Alfaro, S. Fubini and G. Furlan, \textit{"A new classical solution of the Yang-Mills field equations"}, \href{https://www.sciencedirect.com/science/article/abs/pii/0370269376900228?via%3Dihub}{Phys. Lett. B 65, 163 (1976)}.

\bibitem{Corrigan-Fairlie1977} E. Corrigan and D. B. Fairlie, \textit{"Scalar field theory and exact solutions to a classical SU(2) gauge theory"}, \href{https://www.sciencedirect.com/science/article/abs/pii/0370269377908085?via%3Dihub}{Phys. Lett. B 67, 69 (1977)}.

\bibitem{Wilczek1976} F. Wilczek, \textit{"Geometry and interactions of instantons"}, \href{http://inspirehep.net/record/4030?ln=en}{In Quark Confinement and Field Theory, edited by D. Stump and D. Weingarten, (Wiley, New York), pp. 211-219 (1977)}.

\bibitem{Cervero1977} J. Cervero, L. Jacobs and C. R. Nohl, \textit{"Elliptic solutions of classical Yang-Mills theory"}, \href{https://www.sciencedirect.com/science/article/abs/pii/0370269377905640}{Phys. Lett. B 69, 351 (1977)}.

\bibitem{Justin2011} J. Zinn-Justin, \textit{"Barrier penetration and instantons"}, \href{http://landau.gitlab.io/qm/spring16/seminar-6-tunnelirovaniye-i-instantony-v-kvantovoy-mekhanike/zinn-justin.pdf}{Lecture notes, 114 pp. (2011)}.

\bibitem{Gross-Neveu1974} D. J. Gross and A. Neveu, \textit{"Dynamical symmetry breaking in asymptotically free field theories"}, \href{https://journals.aps.org/prd/abstract/10.1103/PhysRevD.10.3235}{Phys. Rev. D 10, 3235 (1974)}.

\bibitem{Minwalla-Yokoyama2015} Sh. Minwalla and Sh. Yokoyama, \textit{"Chern Simons Bosonization along RG Flows"}, JHEP 1602, 103 (2016), \href{http://arxiv.org/abs/1507.04546}{[arXiv:1507.04546 [hep-th]]}.

\bibitem{Girardello2002} L. Girardello, M. Porrati and A. Zaffaroni, \textit{"3-D interacting CFTs and generalized Higgs phenomenon in higher spin theories on AdS"}, Phys. Lett. B 561, 289 (2003), \href{http://arxiv.org/abs/hep-th/0212181}{[arXiv:hep-th/0212181]}.

\bibitem{Petkou2003} A. C. Petkou, \textit{"Evaluating the AdS dual of the critical $O(N)$ vector model"}, JHEP 0303, 049 (2003), \href{http://arxiv.org/abs/hep-th/0302063}{[arXiv:hep-th/0302063]}.

\bibitem{Leigh-Petkou2003} R. G. Leigh and A. C. Petkou, \textit{"Holography of the $\mathcal{N}=1$ higher-spin theory on $AdS_4$"}, JHEP 0306, 011 (2003), \href{http://arxiv.org/abs/hep-th/0304217}{[arXiv:hep-th/0304217]}.

\bibitem{Sezgin-Sundell2003} E. Sezgin and P. Sundell, \textit{"Holography in 4D (super) higher spin theories and a test via cubic scalar couplings"}, JHEP 0507, 044 (2005), \href{http://arxiv.org/abs/hep-th/0305040}{[arXiv:hep-th/0305040]}.

\bibitem{Aharony2011} O. Aharony, G. G.-Ari and R. Yacoby, \textit{"d=3 bosonic vector models coupled to Chern-Simons gauge theories"}, JHEP 1203, 037 (2012), \href{http://arxiv.org/abs/1110.4382}{[arXiv:1110.4382 [hep-th]]}.

\bibitem{Giombi2011} S. Giombi, Sh. Minwalla, Sh. Prakash, S. P. Trivedi, S. R. Wadia and X. Yin, \textit{"Chern-Simons theory with vector fermion matter"}, Eur. Phys. J. C 72, 2112 (2012), \href{http://arxiv.org/abs/1110.4386}{[arXiv:1110.4386 [hep-th]]}.

\bibitem{Hikida-Wada2017} Y. Hikida and T. Wada, \textit{"Marginal deformations of 3d supersymmetric $U(N)$
model and broken higher spin symmetry"}, JHEP 1703, 047 (2017), \href{http://arxiv.org/abs/1701.03563}{[arXiv:1701.03563 [hep-th]]}.

\bibitem{Sezgin-Sundell2005-01} E. Sezgin and P. Sundell, \textit{"An exact solution of 4D higher-spin gauge theory"}, Nucl. Phys. B 762, 1 (2007), \href{http://arxiv.org/abs/hep-th/0508158}{[arXiv:hep-th/0508158]}.

\bibitem{Sezgin-Sundell2005-02} E. Sezgin and P. Sundell, \textit{"On an exact cosmological solution of higher spin gauge theory"}, Bulg. J. Phys. 33, 506 (2006), \href{http://arxiv.org/abs/hep-th/0511296}{[arXiv:hep-th/0511296]}.

\bibitem{Sezgin-Sundell2012} E. Sezgin and P. Sundell, \textit{"Supersymmetric higher spin theories"}, J. Phys. A 46, 214022 (2013), \href{http://arxiv.org/abs/1208.6019}{[arXiv:1208.6019 [hep-th]]}.

\bibitem{Iazeolla-Raeymaekers2015} C. Iazeolla and J. Raeymaekers, \textit{"On big crunch solutions in Prokushkin-Vasiliev theory"}, JHEP 1601, 177 (2016), \href{http://arxiv.org/abs/1510.08835}{[arXiv:1510.08835 [hep-th]]}.

\bibitem{Vasiliev2012} M. A. Vasiliev, \textit{"Holography, unfolding and higher-spin theory"}, J. Phys. A 46, 214013 (2013), \href{http://arxiv.org/abs/1203.5554}{[arXiv:1203.5554 [hep-th]]}.

\bibitem{ABJ} O. Aharony, O. Bergman, D. L. Jafferis, \textit{"Fractional M2-branes"}, JHEP 0811, 043 (2008), \href{http://arxiv.org/abs/0807.4924}{[arXiv:0807.4924 [hep-th]]}.

\bibitem{Gaiotto-Yin2007} D. Gaiotto and X. Yin, \textit{"Notes on superconformal Chern-Simons-Matter theories"}, JHEP 0708, 056 (2007), \href{http://arxiv.org/abs/0704.3740}{[arXiv:0704.3740 [hep-th]]}.

\bibitem{Kim2018} K. K. Kim and O-K. Kwon, \textit{"Janus ABJM models with mass deformation"}, JHEP 1808, 082 (2018), \href{http://arxiv.org/abs/1806.06963}{[arXiv:1806.06963 [hep-th]]}.

\bibitem{Gauntlett2018} J. P. Gauntlett and C. Rosen, \textit{"Susy Q and spatially modulated deformations of ABJM theory"}, JHEP 1810, 066 (2018), \href{http://arxiv.org/abs/1808.02488}{[arXiv:1808.02488 [hep-th]]}.

\bibitem{Arav2019} I. Arav, J. P. Gauntlett, M. Roberts and C. Rosen, \textit{"Spatially modulated and supersymmetric deformations of ABJM theory"}, JHEP 1904, 099 (2019), \href{http://arxiv.org/abs/1812.11159}{[arXiv:1812.11159 [hep-th]]}.

\bibitem{Kim2019} K. K. Kim, Y. Kim, O-K. Kwon and C. Kim, \textit{"Aspects of massive ABJM models with inhomogeneous mass parameters"}, JHEP 1912, 153 (2019), \href{http://arxiv.org/abs/1910.05044}{[arXiv:1910.05044 [hep-th]]}.

\bibitem{Donos2017} A. Donos, J. P. Gauntlett, C. Rosen and O. S.-Rodriguez,\textit{"Boomerang RG flows in M-theory with intermediate scaling"}, JHEP 1707, 128 (2017), \href{http://arxiv.org/abs/1705.03000}{[arXiv:1705.03000 [hep-th]]}.

\bibitem{deHaroSolodukhinSkenderis} S. de Haro, S. N. Solodukhin and K. Skenderis, \textit{"Holographic reconstruction of spacetime and renormalization in the AdS/CFT correspondence"}, Commun. Math. Phys. 217, 595 (2001), \href{http://arxiv.org/abs/hep-th/0002230}{[arXiv:hep-th/0002230]}.

\bibitem{Bianchi3} M. Bianchi, D. Z. Freedman and K. Skenderis,\textit{"Holographic renormalization"}, Nucl. Phys. B 631, 159 (2002), \href{http://arxiv.org/abs/hep-th/0112119}{[arXiv:hep-th/0112119]}.

\bibitem{OoguriVafa016} H. Ooguri and C. Vafa, \textit{"Non-supersymmetric AdS and the Swampland"}, Adv. Theor. Math. Phys. 21, 1787 (2017), \href{http://arxiv.org/abs/1610.01533}{[arXiv:1610.01533 [hep-th]]}.

\bibitem{Freivogel-Kleban2016} B. Freivogel and M. Kleban, \textit{"Vacua Morghulis"}, \href{http://arxiv.org/abs/1610.04564}{[arXiv:1610.04564 [hep-th]]}.

\bibitem{Amsel2007} A. J. Amsel, T. Hertog, S. Hollands and D. Marolf, \textit{"A tale of two superpotentials: Stability and instability in designer gravity"}, Phys. Rev. D 75, 084008 (2007), \href{http://arxiv.org/abs/hep-th/0701038}{[arXiv:hep-th/0701038]}, \href{https://link.aps.org/doi/10.1103/PhysRevD.77.049903} {Erratum: Phys. Rev. D 77, 049903 (2008)}.

\bibitem{Faulkner2010} T. Faulkner, G. T. Horowitz and M. M. Roberts, \textit{"New stability results for Einstein scalar gravity"}, Class. Quant. Grav. 27, 205007 (2010), \href{http://arxiv.org/abs/1006.2387}{[arXiv:1006.2387 [hep-th]]}.

\bibitem{LucaVecchi} L. Vecchi, \textit{"Multitrace deformations, Gamow states, and stability of AdS/CFT"}, JHEP 1104, 056 (2011), \href{http://arxiv.org/abs/1005.4921}{[arXiv:1005.4921 [hep-th]]}.

\bibitem{Giombi2019} S. Giombi, R. Huang, I. R. Klebanov, S. S. Pufu and G. Tarnopolsky, \textit{"The $O(N)$ model in $4<d<6$: Instantons and complex CFTs"}, Phys. Rev. D 101, 045013 (2020), \href{http://arxiv.org/abs/1910.02462}{[arXiv:1910.02462 [hep-th]]}.

\end{thebibliography}
\end{document}